\documentclass{amsart}

\usepackage{amsmath, latexsym, amsthm, amsfonts,bm,amssymb} 
\usepackage{graphicx} 
\usepackage{epsfig}
\usepackage{varioref}

\usepackage{textcomp}
\usepackage{color}

\usepackage{psfrag}
\usepackage{epstopdf}
\usepackage{url}
\usepackage{float}
\usepackage{enumerate}

\usepackage{natbib} 

\usepackage{subfloat}
\usepackage{subfig}

\theoremstyle{plain}
\newtheorem{thm}{Theorem}

\theoremstyle{remark}

\newtheorem{remark}{Remark}

\def\convd{\stackrel{ d}{\rightarrow}}

\def\bSig\mathbf{\Sigma}

\renewcommand{\hat}{\widehat}

\newcommand{\rR}{\mathbb R}

\newcommand{\rd}{{\rm d}}

\def\convd{\stackrel{\mathcal{D}}{\rightarrow}}

\def\ex{{\rm {\mathbb E\,}}}

\allowdisplaybreaks

\bibpunct{(}{)}{;}{a}{}{,} 

\begin{document}

\author{Itai Dattner}
\address{Department of Statistics,\\
University of Haifa,\\
199 Aba Khoushy Ave.,\\
Mount Carmel,\\
Haifa 3498838,\\
Israel}
\email{idattner@stat.haifa.ac.il}

\author{Shota Gugushvili}
\address{Mathematical Institute\\
Leiden University\\
P.O. Box 9512\\
2300 RA Leiden\\
The Netherlands}
\email{shota.gugushvili@math.leidenuniv.nl}

\title[Parameter estimation in ODEs]{Application of one-step method to parameter estimation in ODE models}



\begin{abstract}
In this paper we study application of Le Cam's one-step method to parameter estimation in ordinary differential equations models. This computationally simple technique can serve as an alternative to numerical evaluation of the popular nonlinear least squares estimator, which typically requires the use of a multi-step iterative algorithm and repetitive numerical integration of the ODE system. The one-step method starts from a preliminary $\sqrt{n}$-consistent estimator of the parameter of interest and next turns it into an asymptotic (as the sample size $n\rightarrow\infty$) equivalent of the least squares estimator through a numerically straightforward procedure. We demonstrate performance of the one-step estimator via extensive simulations and real data examples. The method enables the researcher to obtain both point and interval estimates. The preliminary $\sqrt{n}$-consistent estimator that we use depends on nonparametric smoothing, and we provide a data driven methodology for choosing its tuning parameter and support it by theory. An easy implementation scheme of the one-step method for practical use is pointed out. 
\end{abstract}

\maketitle

\section{Introduction}\label{s:int}
Systems of ordinary differential equations (ODEs in short) are commonly used for the mathematical modeling of the rate of change of dynamic processes (e.g., in mathematical biology, see \citet{edelstein2005mathematical}; in the theory of chemical reaction networks, see \cite{feinberg1979lectures} and \cite{sontag2001structure}; and in biochemistry, see \citet{voit2000computational}). Statistical inference for ODEs is not a trivial task, because numerical evaluation of standard estimators, like the maximum likelihood or the least squares estimators, may be difficult or computationally costly. Therefore, over the last few decades, first in the numerical analysis and mathematical biology literature and lately also in the statistical literature, various alternative, primarily nonparametric smoothing based methods have been proposed in the statistical literature to tackle the problem, see, e.g., \cite{Bellman197126}, \cite{varah1982spline}, \cite{voit1982}, \cite{ramsay2007parameter}, \cite{hooker2009forcing}, \cite{hooker2011parameterizing}, \cite{gugushvili2012sqrt}, \cite{campbell2014anova}, \cite{vujavcic2015time}, \cite{dattner2015model}, \cite{dattner2015}, among others. These techniques typically share the property of being computationally simpler, but often also statistically less efficient than the maximum likelihood or the least squares methods.

The ODE systems we have in mind take the form
\begin{equation}
\label{system}
\begin{cases}
{ x}^{\prime}(t)=F({ x}(t),{ \theta}), \quad t\in[0,1],\\
x(0)=\xi
\end{cases}
\end{equation}
where ${ x}(t)=(x_1(t),\ldots,x_d(t))^{tr}$ is a
$d$-dimensional state variable, ${
\theta}=(\theta_1,\ldots,\theta_p)^{tr}$ denotes a $p$-dimensional parameter, while the column $d$-vector $x(0)=\xi$ defines the initial condition. We define $\eta:=(\xi,\theta)$  and denote the solution to \eqref{system} corresponding to the parameter $\eta$ by~\[x(\eta,t):=(x_1(\eta,t),\ldots,x_d(\eta,t))^{tr}.\] 

Knowledge regarding the system parameters $\xi$ and $\theta$ is of vital importance for the study of a process that \eqref{system} models. Indeed, these parameters affect the qualitative properties of the system, and their knowledge allows one to predict the system behaviour. However, in practice the parameter $\theta$ and possibly also the initial condition $\xi$ are unknown to the researcher. Typically they cannot be measured directly, but have to be inferred from noisy measurements of the process under study.

Let $\eta_0=(\xi_0,\theta_0)$ be the `true' parameter value that governs the underlying process. The common statistical model considered for the noisy measurements of the process at time
instances $t_1,\ldots,t_n$ (not necessarily equally spaced), is the additive measurement error model,
\begin{equation}
\label{obs}
Y_{ij}=x_{i}(\eta_0,t_j)+\epsilon_{ij}, \quad i=1,\ldots,d,j=1,\ldots,n,
\end{equation}
where the random variables $\epsilon_{ij}$ are independent 
measurement errors (not necessarily Gaussian). Based on observation pairs $(t_j,Y_{ij}),$ $i=1,\ldots,d,j=1,\ldots,n,$ the goal is to estimate the parameter $\eta_0$.

A classical approach to parameter estimation for ordinary differential equations is the nonlinear least squares (NLS) method. Its use is based on the observation that the problem at hand in its essence is a nonlinear regression problem, where the regression function $x(\eta,\cdot)$ is defined implicitly as the solution to \eqref{system}. The least squares estimator $\widetilde{\eta}_n=(\widetilde{\xi}_n,\widetilde{\theta}_n)$ of $\eta_0$ is defined as a minimizer of the least squares criterion function $R_n(\cdot),$
\begin{eqnarray}
\label{rn}
\widetilde{\eta}_n&=&(\widetilde{\xi}_n,\widetilde{\theta}_n)=\operatorname{argmin}_{\eta} \sum_{i=1}^d\sum_{j=1}^n (Y_{ij}-x_i(\eta,t_j))^2
\\\nonumber&&=:\operatorname{argmin}_{\eta} R_n(\eta).
\end{eqnarray}
The strongest justification for the use of the least squares estimator lies in its attractive asymptotic properties; see e.g., \cite{jennrich1969asymptotic} and \cite{wu1981asymptotic}.
In most practical applications the  solution $x(\eta,\cdot)$ to \eqref{system} is nonlinear in the parameter $\eta$, and therefore some iterative procedure has to be used to compute the nonlinear least squares estimator. Such procedures require an initial guess for a minimizer $\widetilde{\eta}_n,$ and then proceed by constructing successive approximations to the least squares estimator (in a direction guided by the gradient of the criterion function, when a gradient based optimization method, e.g. the Levenberg-Marquardt method, is used). However, the noisy and nonlinear character of the optimization problem may lead for the procedure to end up in a local minimum of the least squares criterion function, especially when good initial guesses of the parameter values are not available. Furthermore, in most of interesting applications the system \eqref{system} is nonlinear and does not have a closed form solution. In that case at every step of the iterative procedure one has to numerically integrate \eqref{system} (as well as the system of the associated sensitivity equations in order to compute the gradient of the criterion function, in case a gradient-based optimization method is used). Since the number of iterations made until convergence of the algorithm can be ascertained is usually large, in most cases this leads to a computational bottleneck. This is the case especially in mathematical biology and biochemistry, where a highly nonlinear character of dependence of the solution $x(\eta,\cdot)$ on the parameter $\eta$ leads to `stiff' integration problems. For a penetrating discussion of these points see e.g.\ \cite{ramsay2007parameter} and \cite{voit2004decoupling}.

Although NLS algorithms and ODE integration routines are constantly improving, and so is the available computational power, admittedly much time and effort can be saved with alternative, less computationally intense approaches, see \cite{voit2004decoupling}. In this paper we explore application of Le Cam's one-step estimator (see, e.g., \cite{vanderVaart1998}) to parameter estimation for systems of ordinary differential equations (ODEs). Some examples of similar studies in different areas are \cite{bickel1975one}, \cite{simpson1992one},  \cite{field1994one}, \cite{cai2000efficient}, \cite{delecroix2003efficient}, and  \cite{rieder2012robust}. In particular, our main goal is to show that the one-step method is at least comparable to NLS, first asymptotically, and second in finite samples. We would like to stress the fact that the one-step method is not simply a numerical approximation to an algorithm used for numerical evaluation of NLS: it is an estimation method on its own.

The main contributions of our paper are: (i) Smoothing-based parameter estimation methods for ODE systems can be upgraded to have statistical efficiency of NLS through a computationally simple one-step method. (ii) If one wants to avoid using NLS (as is often the case in the applied literature, see e.g.\ \cite{stein13} and \cite{bucci16}), one can still do this, while not losing statistical efficiency of NLS and computational properties of smoothing-based methods. (iii) We show how to perform smoothing in a data-driven manner, and provide theory supporting our data-driven algorithm. (iv) We point out a very simple scheme for implementing the one-step estimator, which is readily available in any software that implements Newton-type optimisation algorithms, such as \cite{R} and \cite{matlab}. 

Pertaining to point (i) above, we highlight the extent of loss of efficiency of smoothing-based methods compared to the NLS and the one-step method, which in some simulation setups is of alarming degree. With high throughput, dense-in-time data, that is becoming increasingly available in practice, specifically in molecular biology (see \cite{voit2004decoupling} and \cite{goel2008}), and that would allow an in-depth study of underlying biological processes, such a statistical efficiency loss is clearly undesirable. On the other hand, current ODE inference algorithms must also meet challenges with massive amounts of data and complex models awaiting in the near future. Pertaining to point (ii), as noted in \cite{chou2009recent}, that far no parameter estimation technique for ODEs has arisen as a clear winner in terms of efficiency, robustness and reliability in realistic data scenarios. In this sense, addition of the one-step method (that shares some of the better properties of both the smoothing-based methods and NLS) to a practitioner's toolbox appears a sensible option. Concerning (iii), we note that much of the literature dealing with smoothing-based inference methods for ODEs in practice does smoothing either in a theoretically suboptimal or even an ad hoc way. A distinct advantage of our proposed approach is providing theoretical guarantees for data-depending smoothing that our procedure employs as an intermediate step. Finally, concerning our contribution (iv), we point out an important relation between the one-step estimator and the Levenberg-Marquardt algorithm, which leads to a very practical and straightforward implementation of the method: when computational time is an issue, our simulations and theory justify the use of the Levenberg-Marquardt method with {\it one iteration}, provided it is initialised at an appropriate smoothing-based parameter estimator, since this reduces to the one-step estimation framework. 

The rest of the paper is organized as follows: in Section~\ref{accel} we describe the one-step estimator in the context of ODEs. In Section~\ref{theo} we provide theoretical results for it. Section~\ref{sim} presents a detailed simulation study illustrating the performance of the one-step method, with further examples in Section~\ref{additional}, while Section~ \ref{real} contains numerical results based on  real data examples. 
Section~\ref{sum} summarizes our contribution and outlines potential future research directions. Finally, Appendices give a proof of our theoretical result, and some further implementational details on the methods in the main text of the paper.


%
%
\section{One-step estimate for ODEs}\label{accel}

When one adopts an asymptotic point of view on statistics, all the estimators with the same asymptotic variance can be considered as equivalent. We now demonstrate how once a preliminary $\sqrt{n}$-consistent estimator $\hat{\eta}_n$ of the parameter $\eta$ is available (see below for our choice), one can obtain an asymptotically equivalent estimator to the least squares estimator in just one extra step, referred to as the one-step method in the statistical literature, see e.g.\ Section 5.7 in \cite{vanderVaart1998} for the motivation behind it and a detailed exposition. 

Introduce the function
\begin{equation}
\label{esteq}
\Psi_n(\eta)=\sum_{j=1}^n \psi_{\eta}(t_j,Y_j),
\end{equation}
where
\begin{equation}
\label{psieta}
\psi_{\eta}(t,y)=(x_{\eta}^{\prime}(\eta,t))^{tr}(y-x(\eta,t)),
\end{equation}
with $x_{\eta}^{\prime}(\eta,t)$ denoting the derivative of $x(\eta,t)$ with respect to $\eta.$ Specifically, the $i$th row of $x_{\eta}^{\prime}(\eta,t)$ is the gradient of $x_i(\eta,t)$ with respect to $\eta.$

The one-step estimator $\overline{\eta}_n$ of $\eta_0$ is defined as a solution in $\eta$ of the equation
\begin{equation*}
\Psi_n(\hat{\eta}_n)+\frac{d}{d\eta}\Psi_n(\hat{\eta}_n)(\eta-\hat{\eta}_n)=0.
\end{equation*}
If $\frac{d}{d\eta}\Psi_n(\hat{\eta}_n)$ is invertible, the estimator $\overline{\eta}_n$ can be expressed as
\begin{equation}
\label{opt_par}
\overline{\eta}_n=\hat{\eta}_n-\left( \frac{d}{d\eta}\Psi_n(\hat{\eta}_n) \right)^{-1}\Psi_n(\hat{\eta}_n).
\end{equation}
In order to implement the estimator just defined, the two essential steps that have to be done are i) evaluation of a preliminary estimator $\hat{\eta}_n,$ and ii) evaluation of $\Psi_n(\hat{\eta}_n)$ and the derivative matrix $\frac{d}{d\eta}\Psi_n(\hat{\eta}_n).$ The computational cost for that is very modest. Indeed, as mentioned in Section \ref{s:int}, step i) is very fast, when a smoothing based estimator is used, see examples below. Furthermore, step ii) reduces to requiring just one numerical integration of the sensitivity and variational equations associated with the system \eqref{system}, as we will now explain. This material is standard in the numerical analysis and ODE literature (cf.\ \cite{schittkowski2002} and \cite{ramsay2017}), but perhaps less familiar to statisticians, hence our decison to provide full details. It is helpful to think of $F$ in \eqref{system} as a function of $\eta$ rather than only $\theta$. Thus, we write the right-hand side $F$ of \eqref{system} as $F(x(\eta,t),\eta)$. Differentiating both sides of \eqref{system} with respect to $\eta$ and interchanging the order of a $t$-derivative with an $\eta$-derivative, we get
\begin{equation}
\label{nomer1}
\begin{cases}
\frac{d}{dt}\frac{\partial}{\partial\eta}{ x}(\eta,t)=F_x^{\prime}({ x}(\eta,t),{ \eta})\frac{\partial}{\partial\eta}x(\eta,t)+F_{\eta}^{\prime}(x(\eta,t),\eta),\\
\frac{\partial}{\partial\eta}x (\eta,0)=(1,0)^{tr},
\end{cases}
\end{equation}
where $1$ and $0$ in the initial conditions here and in equations \eqref{nomer2}--\eqref{nomer3} below should be understood as vectors of $1'$s and $0'$s of the appropriate dimensions. The system \eqref{nomer1} is a matrix differential equation and is usually referred to in the literature as a system of sensitivity equations. By replacing $\eta$ with $\hat\eta_n$ we arrive at the system
\begin{equation}
\label{nomer2}
\begin{cases}
\frac{d}{dt}s(t)=F_x^{\prime}({ x}(\hat{\eta}_n,t),{ \hat{\eta}_n})s(t)+F_{\eta}^{\prime}(x(\hat{\eta}_n,t),\hat{\eta}_n),\\
s(0)=(1,0)^{tr},
\end{cases}
\end{equation}
where we have defined $s(t):=\frac{d}{d\eta}{ x}(\hat{\eta}_n,t)$.
Observe that $x(\hat{\eta}_n,\cdot)$ is a known function, because it can be found by integrating \eqref{system} for parameter values $\hat{\xi}_n$ and $\hat{\theta}_n.$ Consequently, the system of sensitivity equations is a linear system with time-dependent coefficients, and hence is relatively straightforward to integrate. 

By differentiating \eqref{nomer1} one more time with respect to $\eta$ and replacing $\eta$ with $\hat{\eta}_n$ we arrive at the following set of variational equations (sometimes called second-order sensitivity equations):
\begin{equation}
\label{nomer3}
\begin{cases}
\frac{d}{dt}z(t)=F_{\eta\eta}^{\prime\prime}( x(\hat{\eta}_n,t),\hat{\eta}_n) )+F_{\eta x}^{\prime\prime} (x(\hat{\eta}_n,t),\hat{\eta}_n) s(t)\\
+\left\{ F_{x\eta}^{\prime\prime} (x(\hat{\eta}_n,t),\hat{\eta}_n) + F_{xx}^{\prime\prime} (x(\hat{\eta}_n,t),\hat{\eta}_n) s(t) \right\}s(t)\\
+F_{x}^{\prime}( x(\hat{\eta}_n,t),\hat{\eta}_n )z(t),\\
z(0)=0,
\end{cases}
\end{equation}
where we have set $z(t):=\frac{\partial^2}{\partial \eta^2}x(\eta,t)$. For each $z_i,$ $i=1,\ldots,d$, the system \eqref{nomer3} is a matrix differential equation and again is a linear system with time-varying coefficients. Here also we can treat $x$ and $s$ as known, for they can be obtained through numerical integration of \eqref{system} and \eqref{nomer2}. The process of obtaining variational equations can be made automatic through a software implementation.

Integration of \eqref{system}, \eqref{nomer2} and \eqref{nomer3} for the parameter value $\hat{\eta}_n$ allows us to compute $\Psi_n(\hat{\eta}_n)$ and $\frac{d}{d\eta} \Psi_n(\hat{\eta}_n)$, and consequently, the one-step estimator $\overline{\eta}_n$. Note that numerical integration of the variational equations (or at least the sensitivity equations) is usually required when computing the least squares estimator via gradient-based optimization methods (unless the gradient is available analytically). However, in our approach we need to do this only \emph{once}.

\begin{remark}
A seemingly more general non-autonomous system than the autonomous system \eqref{system},
\begin{equation*}
\bigg\{
\begin{array}{l}
\tilde{x}^{\prime}(t)=F(\tilde{x}(t),t,\theta),\ t\in[0,1],
\\
\tilde{x}(0)=\tilde{\xi},
\end{array}
\end{equation*}
may and will be reduced to \eqref{system} by a simple substitution $x(t)=(\tilde{x}^{tr}(t), t)^{tr},\ t\in[0,1]$, and $\xi=(\tilde{\xi}^{tr},0)^{tr}.$
\end{remark}

\section{Theory for the one-step method}\label{theo}
The one-step estimation methodology described in the previous section requires the user to first obtain a preliminary $\sqrt{n}$-consistent estimator of parameter of interest. Obviously, one would like such an estimator to be cheap in computational cost. In the context of ODEs, such preliminary estimators were suggested in \cite{Bellman197126} and \cite{varah1982spline}, who use nonparametric smoothing techniques to bypass numerical integration of the ODEs required in evaluation of the maximum likelihood or the least squares estimators. This approach was studied rigorously from the theoretical point of view in \cite{gugushvili2012sqrt} (other relevant references are, e.g., \cite{brunel2008parameter}, \cite{vujavcic2015time} and \cite{dattner2015}). As mentioned, such methods use nonparametric smoothing techniques, and therefore, their good performance crucially depends on an appropriate choice of a `tuning parameter', such as the bandwidth in the case of kernel smoothing, or the number of basis functions in the case of splines. Moreover, this dependence on the bandwidth choice propagates to performance of the one-step estimator. In this section we describe one of the possible preliminary estimators,  provide a data driven scheme for the choice of the tuning parameter, and derive the relevant theory for the one-step method. 

The preliminary estimation works as follows. The observations are first smoothed, which results in an estimator $\hat{x}_n(\cdot)$ for the solution
$x(\eta_0,\cdot)$  of the system, and by differentiation, in an
estimator $\hat{x}_n^{\prime}(\cdot)$ for
$x^\prime(\eta_0,\cdot)$ . Then the estimator for $\theta_0$ is defined as
the minimizer $\hat{\theta}_n$ over $\theta$ of the
 function
\begin{equation}\label{sme}
\int_0^1\parallel\hat{x}_n^{\prime}(t)-F(\hat{x}_n(t);\theta)\parallel^2
w(t)\, \rd t,
\end{equation}
where $w$ is an appropriate weight function, and $\parallel \cdot \parallel$ denotes the standard Euclidean norm. Hence, this approach bypasses the need to integrate the system numerically, and as a result the parameter estimates can be computed extremely quickly, especially when $F$ in \eqref{system} is linear in $\theta.$ Under regularity conditions
\cite{gugushvili2012sqrt} show that this {\it smooth and match estimator}
(SME) $\hat{\theta}_n$ has the $\sqrt{n}$-rate of convergence to
$\theta$. By the general statistical theory, the $\sqrt{n}$-rate of convergence is in fact the best rate one can expect in the present context. This result thus puts the smooth and match method on a solid theoretical ground.

Note that execution of this method does not require the knowledge of the initial values in \eqref{system}. However, it cannot be used to estimate them. If estimation of initial values is of interest, then once the estimator $\hat{\theta}_n$ is at hand, one may obtain an estimator $\hat\xi_n$ by minimizing with respect to $\xi$ the criterion
\begin{equation*}\label{smex0}
\int_0^1\parallel\hat{x}_n(t)-\xi-\int_0^tF(\hat{x}_n(s);\hat{\theta}_n)\rd s\parallel^2
\rd t. 
\end{equation*}
Notice that this is a linear least squares optimization problem and hence is easy to execute. 

Actually, approaches as above are criticized for not being {\it statistically efficient}. In informal terms this means that the resulting estimators do not squeeze as much information out of the data as the least squares estimator does. In more formal terms, their asymptotic variance is larger than that of the least squares estimator. Hence, sometimes it is suggested  (see, e.g., \cite{swartz1975discussion} for an early reference) to use this method only for generating preliminary estimates that should be used later as initial guesses for more accurate methods. Thus, the SME described above is a natural candidate for serving as a preliminary estimator to be used by the one-step method. Now we describe our data driven methodology for choosing the tuning parameter.

Let $\hat{\eta}_{\rho_n}$ denote an estimator of the ODE parameter $\eta_0,$ that depends on smoothing parameter $\rho_n$ (we make the dependence on the sample size $n$ explicit in our notation). As one specific example, $\hat{\eta}_{\rho_n}$ may be a smooth-and-match or an integral estimator (see Appendix \ref{appB}), in which case $\rho_n$ is the bandwidth $h_n.$ Alternatively, $\rho_n$ may also stand for the number of basis functions. Now consider two sequences of positive numbers $\underline{R}_n\leq\overline{R}_n,$ that for every $n$ define an interval $\mathcal{R}_n=[\underline{R}_n,\overline{R}_n].$ This will be an interval in which a user selects his smoothing parameter (in a data-dependent way), when the sample size is equal to $n.$ More specifically, let $N$ be an arbitrary fixed positive integer. For every $n$ consider a grid of size $N$ of smoothing parameters in $\mathcal{R}_n$:

\begin{equation*}
R_n=\{ \rho_n(k)\in\mathcal{R}_n,k=1,\ldots,N \}.
\end{equation*}
Here $k$ indexes smoothing parameter values contained in the candidate set  $R_n$ of smoothing parameter values available to a user.

Now, a data driven one-step estimator can be defined through the following procedure:
\begin{enumerate}
\item Compute $N$ preliminary estimators $\hat{\eta}_{\rho_n(k)}$ for $\rho_n(k)\in R_n.$
\item Compute $N$ one-step estimators $\overline{\eta}_n=\overline{\eta}_n(\hat{\eta}_{\rho_n(k)}).$
\item Set
\begin{equation}
\overline{\eta}_n^{\ast}=\operatorname{argmin}_{ \overline{\eta}_n(\hat{\eta}_{\rho_n(k)}) } \sum_{i=1}^d\sum_{j=1}^n(Y_{ij}-x_i(\overline{\eta}(\hat{\eta}_{\rho_n(k)}),t_j))^2.
\end{equation}
\end{enumerate}
In the simulation study in the next section we demonstrate that this procedure results in an excellent practical performance of the estimator $\overline{\eta}_n^{\ast}.$ In the theorem below we show that it has a sound theoretical basis as well.
\begin{thm}
\label{thm_accel_smoothing}
Assume that the following conditions hold true:
\begin{enumerate}
\item Observation times $t_1,\ldots,t_n$ are i.i.d.\ with a distribution function $F_T$ supported on the interval $[0,T].$
\item Measurement errors $\epsilon_{ij}$'s are i.i.d.\ with mean zero and variance $\sigma^2>0,$ that are also independent of observation times $t_j$'s.
\item The parameter set $H$ is a compact subset of $\mathbb{R}^{d+p}.$
\item For all $\eta\in H $ and $t\in[0,T],$ the third partial derivatives $x_{\eta_j\eta_k\eta_l}^{\prime\prime\prime}(\eta,t)$ of the ODE solution $x({\eta},t)$ exist and are continuous functions of $\eta$ and $t.$
\item The matrix
\begin{equation}
\label{matrix}
I(\eta)=\frac{1}{\sigma^2}\sum_{i=1}^d \int_0^T\Big(\frac{d}{d\eta}{ x_i}(\eta,t)\Big)^{tr}\Big(\frac{d}{d\eta}{ x_i}(\eta,t)\Big) \rd F_T(t)
\end{equation}
is nondegenerate.
\item For every choice of a deterministic sequence of smoothing parameters $\rho_n\in\mathcal{R}_n,$ the resulting estimator $\hat{\eta}_{\rho_n}$ is $\sqrt{n}$-consistent.
\end{enumerate}
Then
\begin{equation}
\label{onestep}
\sqrt{n}(\overline{\eta}_n^{\ast}-\eta_0) \convd  \mathcal{N}\left(0, I(\eta_0)^{-1} \right),
\end{equation}
where $\convd$ denotes convergence in distribution.
\end{thm}

\begin{remark}
\label{efficiency_rem}
Under conditions of Theorem \ref{thm_accel_smoothing}, the limit covariance matrix in \eqref{onestep} coincides with the limit covariance matrix of the least squares estimator; cf.\ Example 5.27 in \cite{vanderVaart1998}.
\end{remark}

\begin{remark}
\label{consistency_sme_rem}
For a smooth-and-match or an integral estimator, $\sqrt{n}$-consistency for any deterministic choice of the bandwidth $h_n\in R_n$ can be achieved, e.g., by taking $\underline{R}_n=\underline{c}n^{-\underline{r}},$ $\overline{R}_n=\overline{c}n^{-\overline{r}},$ for suitably chosen constants $\underline{c},\overline{c},\overline{r},\underline{r}>0.$ Certain freedom in their choice is in fact allowed. As a specific example, the theoretical analysis of \cite{dattner2015} shows that in order to have the $\sqrt{n}$-rate for the integral estimator, one should take a bandwidth $b=O(n^{-1/3})$. Thus, in our practical implementation in subsequent sections we set $B=n^{-1/3}\times(c_1,...,c_N)$, where the $c_j$'s depend on the grid of points on which we evaluate the kernel estimator. 
\end{remark}

\begin{remark}
\label{gauss_newton}
The one-step method as described in Section \ref{accel} requires evaluation of the second derivative $x_{\eta\eta}^{\prime\prime}(\eta,t)$ of the ODE solution $x(\eta,t)$ as part of evaluation of the matrix $\frac{d}{d\eta}\Psi_n(\hat{\eta}_n).$ A standard argument, cf.\ pp.~71--72 in \cite{vanderVaart1998} shows, however, that Theorem \ref{thm_accel_smoothing} still holds true if in the definition of the one-step estimator $\bar{\eta}_n$ in formula \eqref{opt_par}, the matrix $\frac{d}{d\eta}\Psi_n(\hat{\eta}_n)$ is replaced by the matrix
\begin{equation}
\label{approxhessian}
-\sum_{j=1}^n (x_{\eta}^{\prime}(\hat{\eta}_n,t_j))^{tr}x_{\eta}^{\prime}(\hat{\eta}_n,t_j).
\end{equation}
This version of the one-step method is useful when large numerical errors or numerical instability are expected when evaluating $x_{\eta\eta}^{\prime\prime}(\eta,t)$. A further refinement is to employ damping and to replace the derivative matrix $\frac{d}{d\eta}\Psi_n(\hat{\eta}_n)$ with
\[
-\sum_{j=1}^n (x_{\eta}^{\prime}(\hat{\eta}_n,t_j))^{tr}x_{\eta}^{\prime}(\hat{\eta}_n,t_j)-\lambda_n I,
\]
where $\lambda_n>0$ is a damping parameter and $I$ is an identity matrix of appropriate dimensions. The assumption for the asymptotic theory to go through is that $\lambda_n/n\rightarrow 0$ as $n\rightarrow\infty.$ The idea of this version of the one-step method is that it numerically robustifies the one-step procedure in case the matrix \eqref{approxhessian} is nearly singular (which is not uncommon in practice). We use this version of the one-step method in our simulation example in Section \ref{additional}.
\end{remark}

\subsection{Confidence intervals}
\label{conf_sect}
Clearly, confidence intervals for parameter $\eta_0$ can be generated using equations \eqref{matrix} and \eqref{onestep}. However, the Fisher information matrix in \eqref{matrix} depends on the true values of the parameters, initial values, and $\sigma^2$, which are not known in practice. Fully data driven confidence intervals can be constructed by estimating the Fisher information matrix. To that end we estimate $\sigma^2$ by 
\[
\hat\sigma^2=\frac{1}{d(n-1)}\sum_{i=1}^d\sum_{j=1}^n(Y_{ij}-x_i(\bar\eta_n^*,t_j))^2,
\] 
where $x(\bar\eta_n^*,\cdot)$ stands for the solution of the system \eqref{linear} using the estimated parameters and initial values obtained from the one-step method. 
Then an estimate for the asymptotic variance of the estimator of the parameter $\eta_j$ is given by $I^{-1}_{jj}(\bar\eta_{n}^*)/n$, where $I_{jj}^{-1}(\bar\eta_{n}^*)$ stands for the $j$th diagonal element of the inverse Fisher information matrix evaluated in point $\bar{\eta}_n^*.$ When $s(\cdot)$ has no closed form, the integral in \eqref{matrix} is evaluated using numerical integration (in our examples we will use the trapezoidal rule). Specifically, an approximate $1-\alpha$ level confidence interval for $\eta_{0j}$ is given by
\begin{eqnarray}\label{ci}
[\bar\eta_{j,n}^*-z_{1-\alpha/2} I^{-1/2}_{jj}(\bar\eta_{j,n}^*)/\sqrt{n},\bar\eta_n^*+z_{1-\alpha/2} I^{-1/2}_{jj}(\bar\eta_{j,n}^*)/\sqrt{n}],
\end{eqnarray}
where $z_{1-\alpha/2}$ is the $1-\alpha/2$ quantile of the standard normal distribution.

\section{Simulation study}\label{sim}

In this section we present the results of an extensive simulation study comparing the one-step method to the classical NLS approach. The models we use are standard test examples for parameter inference in ODEs, as indicated in the references we will supply in the relevant places. Our goal is to exhibit that the one-step algorithm provides statistical accuracy comparable to the NLS method in practical scenarios.

All computations in the present section were carried out using Matlab (the code will be sent by the first author upon request). The algorithms we used for computing the NLS and one-step estimators are `default', in the sense that we did not attempt to tweak them to fit better in specific problems. Specifically, the NLS estimator was computed using the Levenberg-Marquardt (\cite{marquardt1963algorithm}) algorithm of Matlab. The variant of SME $\hat{\eta}_n$ that we used in the present and next sections to compute the one-step estimator $\bar{\eta}_n$ is detailed in Appendix B. The local polynomial estimator in some of our examples was based on the implementation from \cite{cao2008matlab}. Further software and hardware details are: Windows 8.1 Pro, Intel \textregistered\ Core\texttrademark\ i7-4550U CPU @ 1.50GHz.

\subsection{Linear ODE}
We start with illustrating the performance of the one-step estimator when used to estimate the parameter and initial value of a one-dimensional linear ordinary differential equation
\begin{equation}
\label{linear}
\begin{cases}
x^{\prime}(t)=\theta_0 x(t),\,\ t\in[0,T], \\
x(0)=\xi_0.
\end{cases}
\end{equation}
This is a toy example, but it allows us to explore the practical performance of the one-step method in great detail and to compare it to the theoretically expected results. Advanced examples will be considered later on.

The solution of the initial value problem \eqref{linear} is $x(t)=x_0\exp(\theta_0 t)$. We generate (pseudo) random observations from the model 
\[Y_j=\xi_0\exp(\theta_0 t_j)+\epsilon_j,\]
where $t_j\in\{0(0.1)10\}$ ($n=101$ ), and $\epsilon_j\sim N(0,0.05^2)$, $j=1\ldots,n$. We consider
\[
\theta_0\in\{-1,-0.8,-0.6,-0.4,-0.2,0.1,0.3,0.5,0.7,0.9\}\]
and $\xi_0\in\{0.5,1\}$. For each pair $(\xi_0,\theta_0)$ we run a Monte Carlo study of $500$ samples of $Y_1\ldots,Y_{101}$, where in each sample we apply both the one-step method and the nonlinear least squares method.  This simulation study enables us to estimate the asymptotic variance of the least squares and the one-step methods. We then compare the results to the true asymptotic variance. The true and estimated asymptotic variances can be obtained for each set of parameters and initial values by inverting the Fisher information matrix; see Subsection \ref{conf_sect}. The optimal bandwidth $b$ used to compute SME was chosen in the set
\[
n^{-1/3} \times ( 0.02,    0.3511,    0.6822,    1.0134,    1.3445,    1.6756,    2.0067,    2.3378,    2.6689,    3
 ),
\]
using the procedure outlined in Remark \ref{consistency_sme_rem}; cf.\ Theorem \ref{thm_accel_smoothing}. We also note that in order not to overload the paper with reporting various tuning constants that depend on specific experimental setups, we will not indicate $c_1,\ldots,c_N$ from Remark  \ref{consistency_sme_rem} in our subsequent examples, but will supply them to the reader by email, should he want to know them.

\begin{figure}
 \centering
\includegraphics[width=0.48\textwidth]{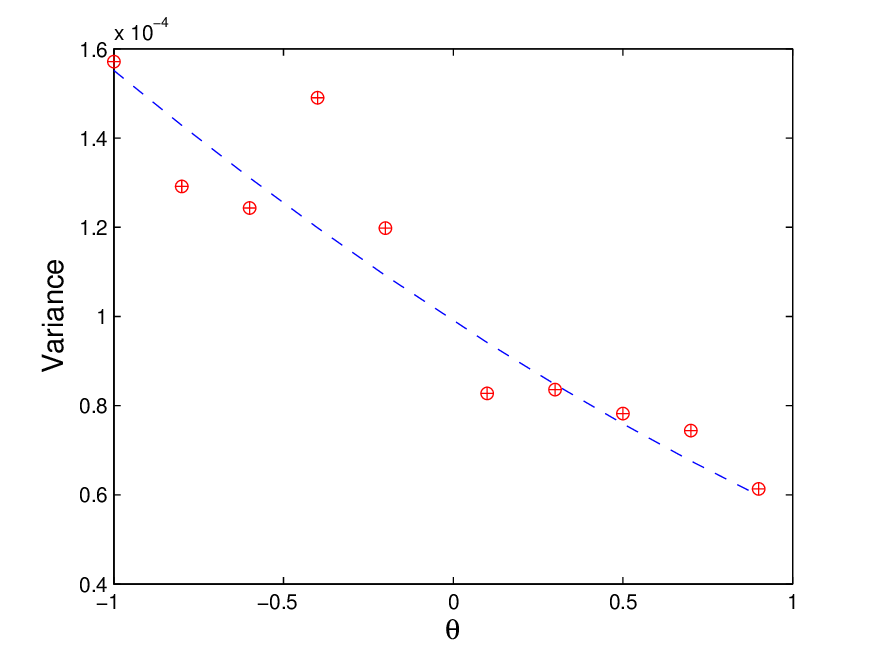}
\includegraphics[width=0.48\textwidth]{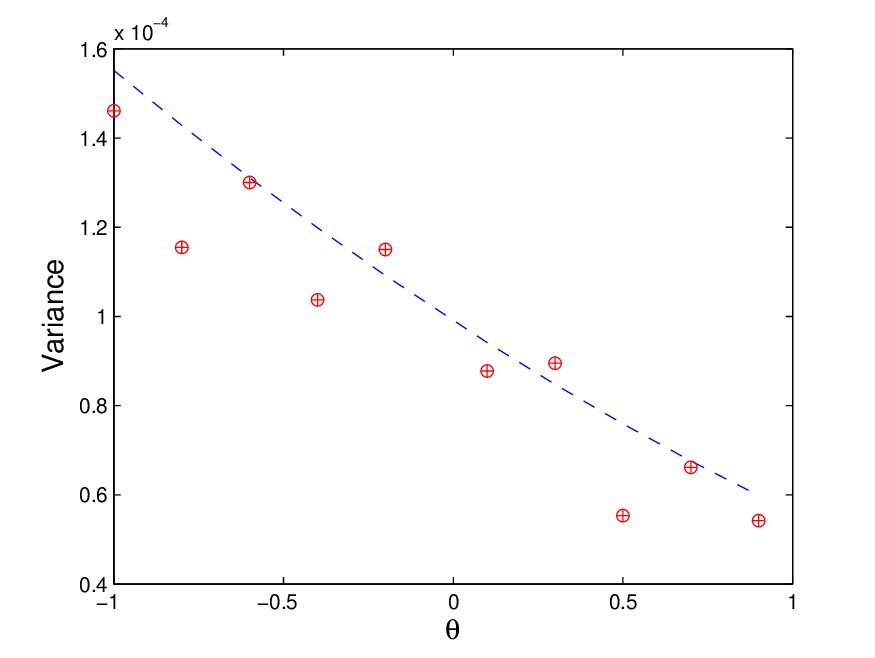}
    \caption{\label{fig:asym_varx0} The estimated variance of the one-step (plus signs) and NLS (circles) estimators $\bar\xi_n$ and $\tilde\xi_n$, respectively, based on $500$ simulations with $n=101$ and $\epsilon_j\sim N(0,0.05^2)$, $j=1\ldots,n$. The estimates are superimposed on the theoretical asymptotic variance (dashed line). The left plot is for $\xi_0=0.5$ and the right one is for $\xi_0=1$. }
\end{figure}

A direct computation gives that in model \eqref{linear} the asymptotic variance of $\bar\xi_n$ depends on $\theta,$ but is independent of the values of $\xi$ itself. In Figure~\ref{fig:asym_varx0} we plot the estimated variance of the one-step estimators (plus signs) and that of the NLS (circles), for estimating $\xi_0$ based on $500$ simulation runs. The estimates are superimposed on the theoretical asymptotic variance (dashed line). The left plot is for $\xi_0=0.5$ and the right one is for $\xi_0=1$. As the theory suggests, independently of the values of $\xi$, the true asymptotic variance is the same. Note that in this specific numerical example the estimated variances of the one-step and NLS estimators are the same. This is not surprising, since in order to apply the NLS we used as the initial point in the parameter space the SME (resulted from using the bandwidth $3\times n^{-1/3}$; this choice was arbitrary). The estimated variances agree with the asymptotic one. We note that the grid of $\theta_0$ does not include $0$, where the asymptotic variance equals zero.   

\begin{figure}
 \centering
\includegraphics[width=0.48\textwidth]{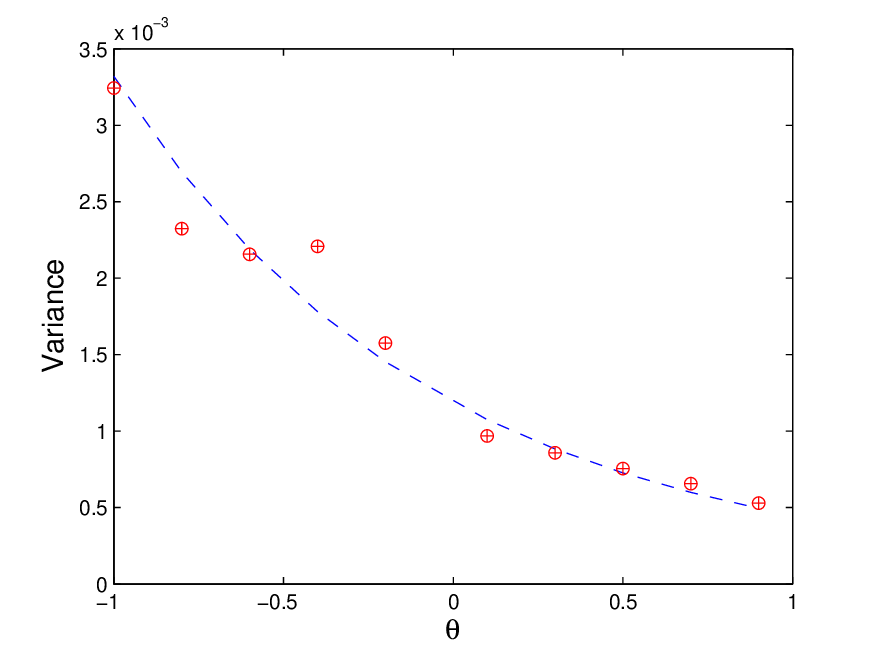}
\includegraphics[width=0.48\textwidth]{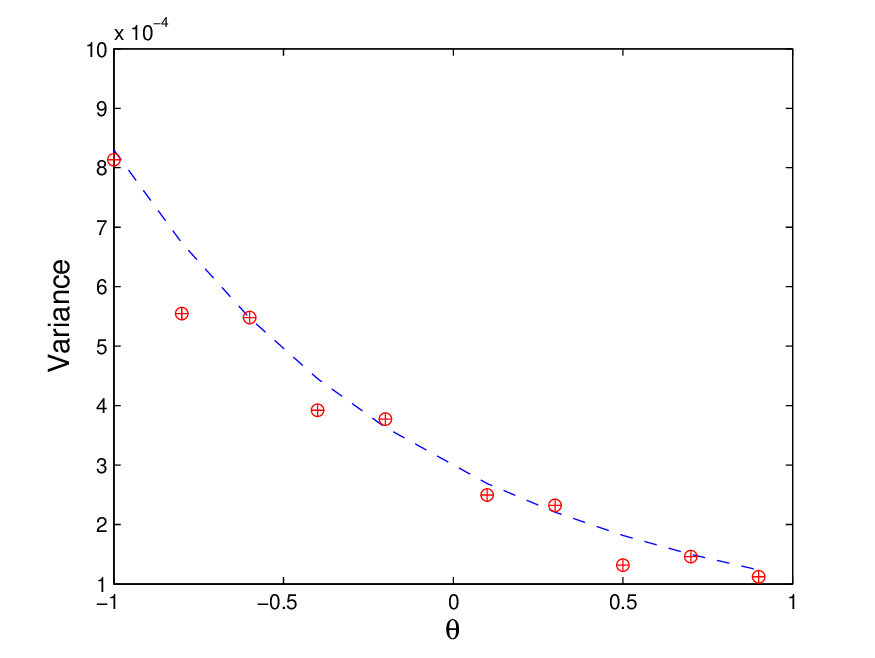}
    \caption{\label{fig:asym_vartheta} The estimated variance of the one-step (plus signs) and NLS (circles) estimators $\bar\theta_n$ and $\tilde\theta_n$, respectively, based on $500$ simulations with $n=101$ and $\epsilon_j\sim N(0,0.05^2)$, $j=1\ldots,n$. The estimates are superimposed on the theoretical asymptotic variance (dashed line). The left plot is for $\xi_0=0.5$ and the  right one is for $\xi_0=1$. }
\end{figure}

In Figure~\ref{fig:asym_vartheta} we see similar plots corresponding to estimating the asymptotic variances of $\bar\theta_n$. Here the variance has different order, depending on the value of $\xi_0$. Again, the estimated variances of the one-step (plus signs) and NLS (circles) estimators are the same, and both agree with the asymptotic one (dashed line). Similar plots were obtained when considering other values for $\sigma^2,$ and therefore we do not present them here. 

In Table~\ref{tab:simple} we present the empirical coverage of various confidence intervals based on a Monte Carlo study with $500$ simulations for different experimental setups. The results should be compared to the nominal coverage of $95$\%. We consider 4 setups denoted by $A,B,C,D$ according to $(\xi_0=1/2,\theta_0=-1), (\xi_0=1/2,\theta_0=1),(\xi_0=1,\theta_0=-1),(\xi_0=1,\theta_0=1)$, respectively. Each scenario is tested for $n=21$, and $n=51$. Table~\ref{tab:simple} presents the point and interval estimates for the parameters of each scenario. We see that the coverage of the confidence intervals is satisfying across the different experimental scenarios. 
\begin{table}
\caption{\label{tab:simple} Means of point estimates and actual coverage of interval estimates for the parameters of model \eqref{linear} according to 4 different experimental setups. The results are based on $500$ simulation runs. The observations are generated according to $Y_j=\xi_0\exp(\theta_0 t_j)+\epsilon_j$, where $t_j\in\{0(0.5)10\}$ ($n=21$ ), or $t_j\in\{0(0.2)10\}$ ($n=51$ ) and $\epsilon_j\sim N(0,0.05^2)$, $j=1\ldots,n$.  The point estimates are given by \eqref{opt_par}; the interval estimates are defined in \eqref{ci}.}																					
\centering
\fbox{%
\begin{tabular}{lccccccc}
\multicolumn{4}{c}{}&
\multicolumn{2}{c}{one-step}&
\multicolumn{2}{c}{NLS}\\
\multicolumn{1}{c}{Setup}&	
\multicolumn{1}{c}{}&	
\multicolumn{1}{c}{}&	
\multicolumn{1}{c}{}&
\multicolumn{1}{c}{Mean}&
\multicolumn{1}{c}{Coverage}&
\multicolumn{1}{c}{Mean}&
\multicolumn{1}{c}{Coverage}\\						
\hline	
n=21&A &$\xi_0$ &   0.500 &    0.501 &     0.942 &   0.501 &    0.942  \\ 
   &&$\theta_0$ &  -1.000 &   -1.002 &     0.946 &  -1.002 &    0.946  \\ 
  &B &$\xi_0$ &   0.500 &    0.500 &     0.928 &   0.500 &    0.928  \\ 
   &&$\theta_0$ &   1.000 &    1.000 &     0.938 &   1.000 &    0.938  \\ 
  &C &$\xi_0$ &   1.000 &    0.999 &     0.932 &   0.999 &    0.932  \\ 
   &&$\theta_0$ &  -1.000 &   -0.997 &     0.940 &  -0.997 &    0.940  \\ 
  &D &$\xi_0$ &   1.000 &    1.000 &     0.944 &   1.000 &    0.944  \\ 
   &&$\theta_0$ &   1.000 &    1.000 &     0.948 &   1.000 &    0.948  \\ 
\hline
 n=51&A &$\xi_0$ &   0.500 &    0.500 &     0.944 &   0.500 &    0.944  \\ 
   &&$\theta_0$ &  -1.000 &   -0.998 &     0.944 &  -0.998 &    0.944  \\ 
  &B &$\xi_0$ &   0.500 &    0.500 &     0.946 &   0.500 &    0.946  \\ 
   &&$\theta_0$ &   1.000 &    0.999 &     0.958 &   0.999 &    0.958  \\ 
  &C &$\xi_0$ &   1.000 &    0.999 &     0.932 &   0.999 &    0.932  \\ 
   &&$\theta_0$ &  -1.000 &   -0.999 &     0.938 &  -1.000 &    0.938  \\ 
  &D &$\xi_0$ &   1.000 &    1.000 &     0.948 &   1.000 &    0.948  \\ 
   &&$\theta_0$ &   1.000 &    1.001 &     0.952 &   1.001 &    0.952  \\ 
\end{tabular}    }
\end{table}

\subsection{Lotka-Volterra system}
The Lotka-Volterra system of ODEs (\cite{edelstein2005mathematical}) is a population dynamics model that describes evolution over time of the populations of two species, predators and their preys. The system takes the form
\begin{equation}\label{lotka}
\bigg\{
\begin{array}{l}
x_1^{\prime}(t)=\theta_1x_1(t)-\theta_2x_1(t)x_2(t),
\\
x_2^{\prime}(t)=-\theta_3x_2(t)+\theta_4x_1(t)x_2(t).
\end{array}
\end{equation}
Here $x_1$ represents the size of the prey population and $x_2$ of
the predator population. In Table \ref{tab:lotka} we see the empirical coverage of the 95\% confidence intervals based on a Monte Carlo study consisting of $500$ simulation runs for different sample sizes.
\begin{table}
\caption{\label{tab:lotka}
Means of point estimates and actual coverage of interval estimates for the parameters of model \eqref{lotka}, where the initial values are $\xi_0=(1,1/2)^{tr}$, and the rate parameters are $\theta_0=(1/2,1/2,1/2,1/2)^{tr}$. The results are based on running $500$ simulations. The observed time points are equidistant on $[0,10]$, and the errors are normal with zero expectation and $\sigma=0.05$. The one-step point estimates are given by \eqref{opt_par}; the interval estimates are defined in \eqref{ci}.}																					
\centering
\fbox{%
\begin{tabular}{lccccccc}
\multicolumn{3}{c}{}&
\multicolumn{2}{c}{one-step}&
\multicolumn{2}{c}{NLS}\\
\multicolumn{1}{c}{Setup}&	
\multicolumn{1}{c}{}&	
\multicolumn{1}{c}{}&
\multicolumn{1}{c}{Mean}&
\multicolumn{1}{c}{Coverage}&
\multicolumn{1}{c}{Mean}&
\multicolumn{1}{c}{Coverage}\\						
\hline	
n=21 &  $\xi_1$ &   1.000 &    1.000 &     0.932 &   0.999 &    0.928  \\ 
 &  $\xi_2$ &   0.500 &    0.500 &     0.936 &   0.500 &    0.934  \\ 
 &  $\theta_1$ &   0.500 &    0.502 &     0.942 &   0.501 &    0.942  \\ 
 &  $\theta_2$ &   0.500 &    0.502 &     0.932 &   0.501 &    0.938  \\ 
 &  $\theta_3$ &   0.500 &    0.500 &     0.910 &   0.501 &    0.916  \\ 
 &  $\theta_4$ &   0.500 &    0.500 &     0.918 &   0.501 &    0.922  \\ 
\hline
n=51 &  $\xi_1$ &   1.000 &    1.000 &     0.958 &   1.000 &    0.966  \\ 
 &  $\xi_2$ &   0.500 &    0.500 &     0.954 &   0.500 &    0.948  \\ 
 &  $\theta_1$ &   0.500 &    0.502 &     0.964 &   0.500 &    0.968  \\ 
 &  $\theta_2$ &   0.500 &    0.501 &     0.968 &   0.500 &    0.964  \\ 
 &  $\theta_3$ &   0.500 &    0.500 &     0.958 &   0.500 &    0.958  \\ 
 &  $\theta_4$ &   0.500 &    0.500 &     0.952 &   0.500 &    0.958  \\ 
\end{tabular}    }
\end{table}

The experimental setup is as follows: the observed time points are equidistant on $[0,10]$; the errors are normal with zero mean and standard deviation $\sigma=0.05$; the initial values are $\xi_0=(1,1/2)^{tr}$, and the parameters are $\theta_0=(1/2,1/2,1/2,1/2)^{tr}$. The point estimates are given by \eqref{opt_par}, while the interval estimates are defined in \eqref{ci}. As expected, the coverage is much better when the sample size is larger. The performance of the one-step and NLS methods is similar.  

In Table \ref{tab:lotka_var} we present the square root of the average of the estimates of the asymptotic variance over the $500$ simulations (denoted by `ASYM'). Next to that we present standard errors of the point estimates as calculated based on the $500$ simulations (denoted by `STE'). The results for both the NLS and one-step methods agree with each other. Note also the first column of this table, where we report the standard errors of the SME, which are larger than those of the one-step, as expected. In this experimental setup the loss of statistical efficiency of SME in comparison to the one-step method and NLS is relatively small, given moderate sample sizes ($n=21$ and $n=51$). See, however, the next subsection.
\begin{table}
\caption{\label{tab:lotka_var}
Standard errors of the point estimates as calculated based on the $500$ simulations (denoted by `STE'). Square root of the average of the estimates of the asymptotic variance, over the $500$ simulations (denoted by `ASYM'). The experimental setup is as in Table \ref{tab:lotka}.}																					
\centering
\fbox{%
\begin{tabular}{lcccccccc}
\multicolumn{2}{c}{}&
\multicolumn{1}{c}{SME}&
\multicolumn{2}{c}{one-step}&
\multicolumn{2}{c}{NLS}\\
\multicolumn{1}{c}{Setup}&	
\multicolumn{1}{c}{}&
\multicolumn{1}{c}{STE}&
\multicolumn{1}{c}{STE}&
\multicolumn{1}{c}{ASYM}&
\multicolumn{1}{c}{STE}&
\multicolumn{1}{c}{ASYM}&\\				
\hline	
n=21 & $\xi_1$ &0.033&    0.025 &    0.023 &     0.025 &   0.023   \\ 
 & $\xi_2$ & 0.022   &0.020 &    0.019 &     0.020 &   0.019   \\ 
 & $\theta_1$ & 0.030 &  0.027 &    0.026 &     0.027 &   0.026   \\ 
 & $\theta_2$ &0.024  &  0.022 &    0.021 &     0.022 &   0.021   \\ 
 & $\theta_3$ & 0.025 &  0.022 &    0.020 &     0.022 &   0.020   \\ 
 & $\theta_4$ & 0.021 &  0.020 &    0.018 &     0.020 &   0.018   \\ 
\hline
n=51 & $\xi_1$ &0.021 &   0.015 &    0.016 &     0.014 &   0.016   \\ 
 & $\xi_2$ & 0.014  & 0.013 &    0.013 &     0.013 &   0.013   \\ 
 & $\theta_1$ &0.019&    0.016 &    0.017 &     0.016 &   0.017   \\ 
 & $\theta_2$ &0.015 &   0.013 &    0.014 &     0.013 &   0.014   \\ 
 & $\theta_3$ & 0.014 &  0.013 &    0.013 &     0.013 &   0.014   \\ 
 & $\theta_4$ & 0.013 &  0.012 &    0.012 &     0.012 &   0.012   \\ 
\end{tabular}    }
\end{table}			
\subsection{Comparison with other methods}
The main theme of this paper is not to compare various parameter estimation methods for ODEs, but to show how a non-efficient estimation method such as SME can be improved statistically, to an efficient one, and to test its practical performance. Indeed, this point was demonstrated above by comparing the variance of the one-step estimator to that of the least squares, which is not considered as a competitor, but serves as a `gold standard' for efficient estimation. For completeness, however, we report results of a small scale comparison that can shed some additional light on the statistical effects of the one step correction on SME. In Table~\ref{tab:comp} we present the results of a simulation study for several experimental setups of the linear ODE case (cf.\ equation \eqref{linear}). The results should be compared to Table~1 of \cite{hall2014quick}, where a different variant of SME is studied. The one-step estimator is uniformly (over all experimental setups) better than the method developed in the aforementioned paper, even though unlike that work we estimate both the initial value and the parameter, and hence have to deal with greater uncertainty. The reduction in standard error achieved by the one-step estimator over the SME is in the range of 30-50\% in this example. Such an improvement of an efficient parameter estimation method over SME is not an isolated instance: \cite{hall2014quick} report results of a Monte Carlo comparison between their version of SME and the generalised smoothing (or profiling) approach of \cite{ramsay2007parameter}, and find out that the latter produces twice as small standard errors for parameter estimates in a specific experimental setup in the FitzHugh-Nagumo model; this despite the fact that the SME in \cite{hall2014quick} relies on a fully observed FitzHugh-Nagumo model, whereas \cite{ramsay2007parameter} assume only one state variable out of two is measured. A lesson to be drawn from this discussion from the statistical efficiency point of view is that one should be very careful when using SME, so as to fully utilise precious information contained in observations.

\begin{table*}
\caption{\label{tab:comp}
The rows in the table correspond (respectively) to the means of point estimates, Monte Carlo empirical standard deviation, means of estimated asymptotic standard deviation, true asymptotic standard deviation, and actual coverage of interval estimates (using the estimated asymptotic standard deviation) for the parameter $\theta_0=1$ in the linear ODE case \eqref{linear} (initial value $\xi_0=1$ was estimated as well). The results are based on $1000$ Monte Carlo simulations. The observed time points are equidistant on $[0,10]$, and the errors are normal with zero expectation and $\sigma$ as in the table. The one-step point estimates are given by \eqref{opt_par}; the interval estimates are defined in \eqref{ci}.}																					
\centering
\fbox{%
\begin{tabular}{ccccccccc}
\multicolumn{3}{c}{n=250}&
\multicolumn{3}{c}{n=500}&
\multicolumn{3}{c}{n=1000}\\	
\multicolumn{1}{c}{$\sigma=0.1$}&	
\multicolumn{1}{c}{$\sigma=0.2$}&
\multicolumn{1}{c}{$\sigma=0.3$}&
\multicolumn{1}{c}{$\sigma=0.1$}&	
\multicolumn{1}{c}{$\sigma=0.2$}&
\multicolumn{1}{c}{$\sigma=0.3$}&
\multicolumn{1}{c}{$\sigma=0.1$}&	
\multicolumn{1}{c}{$\sigma=0.2$}&
\multicolumn{1}{c}{$\sigma=0.3$}\\		
\hline	
1.0010	&	1.0010	&	0.9990	&	1.0000	&	1.0000	&	1.0010	&	1.0000	&	1.0000	&	1.0000	\\
0.0130	&	0.0260	&	0.0400	&	0.0090	&	0.0190	&	0.0280	&	0.0070	&	0.0130	&	0.0200	\\
0.0130	&	0.0280	&	0.0390	&	0.0090	&	0.0180	&	0.0290	&	0.0070	&	0.0130	&	0.0200	\\
0.0130	&	0.0270	&	0.0400	&	0.0100	&	0.0190	&	0.0290	&	0.0070	&	0.0130	&	0.0200	\\
0.9460	&	0.9540	&	0.9610	&	0.9540	&	0.9520	&	0.9530	&	0.9430	&	0.9550	&	0.9470	\\
\end{tabular}    }
\end{table*}				
\subsection{Computational times}
We close this section by reporting one more comparison. Namely, we compare `default' implementations of one-step and NLS with respect to computational time. \cite{voit2004decoupling} consider a test example that was introduced in \cite{robertson1966solution} and point out that it is now frequently used as a
benchmark for the efficiency of stiff solvers. The system is given by 
\begin{equation}\label{alpha1}
\begin{array}{l}
x_1^{\prime}(t)=\theta_1x_2(t)x_3(t)-\theta_2x_1(t),
\\
x_2^{\prime}(t)=\theta_2x_1(t)-\theta_1x_2(t)x_3(t)-\theta_3(x_2(t))^2,
\\
x_3^{\prime}(t)=\theta_3(x_2(t))^2,
\end{array}
\end{equation}
with initial values $\xi_0=(1,0,0)^{tr}$ and parameters $\theta_0=(10^4,0.04,3\times 10^7)^{tr}$. We take the observational time interval to be (in seconds) $[0(0.5)10],$ implying that we have $n=21$ equispaced observations at our disposal. The variance of the noise is set to be $0.01$ times the mean values of the (true) solutions corresponding to the system just defined. The actual coverage of the confidence intervals for the parameters $(\theta_1,\theta_2,\theta_3)^{tr}$ for a nominal level of $95\%$, and using the one-step and NLS estimator based on $100$ Monte Carlo simulations was $(1,0.97,1)^{tr}\times 100\%$. The widths of the confidence intervals for one-step and NLS were comparable. A single evaluation of the one-step estimator took about $26$ seconds on average, while that of the NLS took about $78$ seconds.

However, one should keep in mind that a completely objective comparison of computational costs for various ODE inference techniques is hardly possible, as this depends on factors like software and hardware used, as well as the skill of the user in tailoring the methods to specific applications. Also, one cannot expect that a single best  method (as far as the computational cost is concerned) will emerge accross all possible experimental setups (different ODE systems, sample sizes, time scales and resolutions, noise levels).

\section{Further comparison}
\label{additional}

In this section we additionally study a notoriously difficult test example in parameter inference for ODEs. In particular, we illustrate the fact why it might be advantageous to use the one-step method instead of a `default' implementation of NLS, such as the Levenberg-Marquardt algorithm in Matlab. Our take-home message is that overreliance on `default' implementations of NLS estimation routines for ODEs is perhaps a strategy to be critically reconsidered. We also point out a very simple practical scheme for implementing the one-step method.

\subsection{Goodwin's oscillator}

Goodwin's oscillator, see \cite{goodwin1963}, \cite{goodwin1965} and \cite{griffith1968}, is a simple ODE system for modelling feedback control in gene regulatory mechanisms. Various versions of this model have been used as test examples for MCMC samplers in the Bayesian approach to inference in ODE models, see, e.g., \cite{girolami2008}, \cite{calderhead2009}, \cite{oates2017} and \cite{oates2017b}. Standard Metropolis-Hastings samplers encounters severe difficulties in this setting due to a highly complex shape of the likelihood the Goodwin oscillator typically produces, with Markov chains getting trapped in local maxima of the likelihood surface. Not surprisingly, similar behaviour can be observed also in the case of default implementations of the least squares routines, as we will now demonstrate.

\begin{figure}
 \centering
\includegraphics[width=0.8\textwidth,height=1.8\textheight,keepaspectratio]{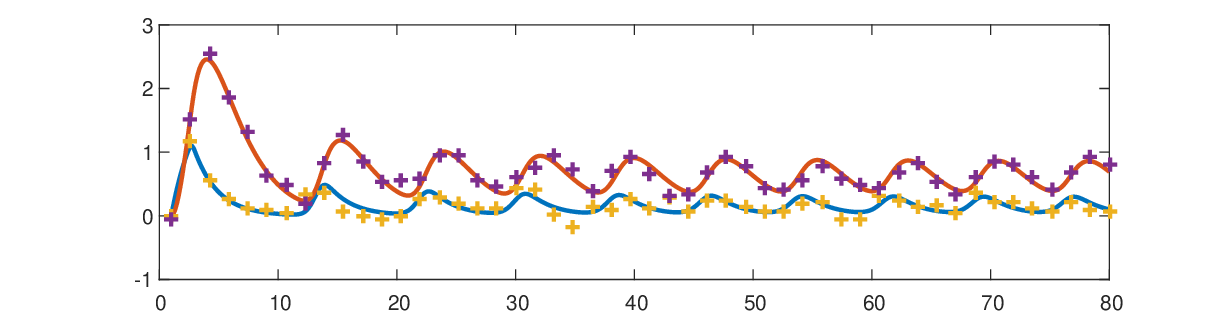}
    \caption{\label{fig:realization} Components of the solution $x_1$ and $x_2$ of the system \eqref{goodwin} (red and blue solid lines) with a typical realisation of noisy observations (purple and yellow crosses).}
\end{figure}

The following version of Goodwin's model is described e.g.\ in \cite{murray2002}, while the experimental setup mimics the one in \cite{oates2017}. The ODE system we consider is
\begin{equation}
\label{goodwin}
\begin{split}
\begin{cases}
{ x}^{\prime}_1(t)&=\frac{\theta_1}{1+\theta_2x_3(t)^{10}}-\theta_5 x_1(t),\\
{ x}^{\prime}_2(t)&=\theta_3 x_1(t)-\theta_5 x_2(t),\\
{ x}^{\prime}_3(t)&=\theta_4 x_2(t)-\theta_5 x_3(t),
\end{cases}
\end{split}
\end{equation}
We used the following parameter values,
\[
\theta_1=1, \theta_2=3, \theta_3=2, \theta_4=1, \theta_5=0.5,
\]
and zero initial conditions. Initial conditions and all the parameters except $\theta_1$ and $\theta_5$ were assumed to be known in the estimation problem. We compare the performance of the NLS and the one-step method through 100 Monte Carlo simulations for estimating the parameter $\theta=(\theta_1, \theta_5)^{tr}$. We consider the case when \eqref{goodwin} is observed only partially, with observations on $x_3$ not available; observed are the variables $x_1,x_2$ subject to additive Gaussian errors, with $n=50$ noisy observations spread uniformly over the time interval  $[0,80]$. The solution to \eqref{goodwin} shows a characteristic oscillatory behaviour, and we plot it in Figure \ref{fig:realization} together with corresponding observations in one simulation run.

We consider three scenarios corresponding to three noise levels $\sigma=0.01,0.15, 0.25$, respectively. It turned out that in this specific example the version of the one-step method that we described in Section \ref{theo} in Remark \ref{gauss_newton} produced better results than the core one-step method from Section \ref{accel}, so that we decided to perform a comparison of this version to a default implementation of the Levenberg-Marquardt method in Matlab. Numerically the one-step method in this case reduces to one iteration of the Levenberg-Marquardt algorithm, but with a difference that it is initialized at the $\sqrt{n}$-consistent preliminary parameter estimator and not an arbitrary initial guess. The default (starting) value for the damping parameter $\lambda$ of the Levenberg-Marquardt algorithm in Matlab is $\lambda=0.01,$ which is also the one we used for the one-step method. Matlab successively increases the damping parameter until a proposed parameter move of the Levenberg-Marquardt method results in a decrease of the criterion function (the total number of proposals in one optimisation run can be controlled by setting the maximal number of function evaluations for the algorithm). This then constitutes one iteration of the Levenberg-Marquardt method in Matlab. 

\begin{figure}
 \centering
\includegraphics[width=0.8\textwidth,height=1.8\textheight,keepaspectratio]{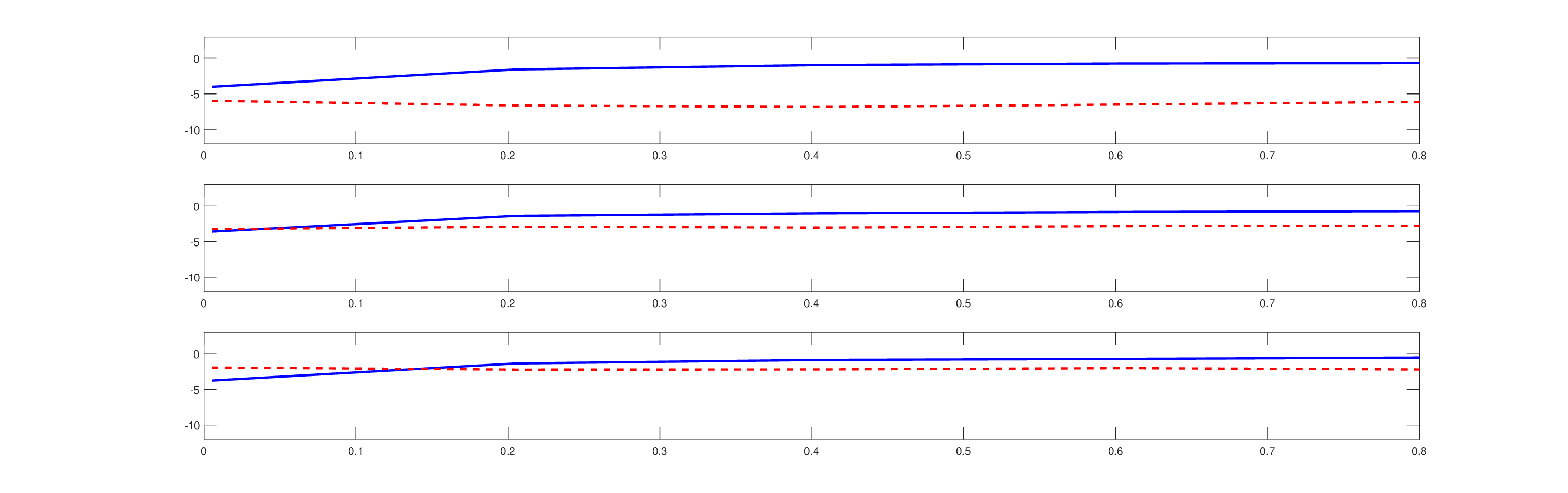}
    \caption{\label{fig:godwinMSE} Simulation results for Goodwin's oscillator in Section \ref{additional}. The $y$-axis gives the logarithm of the sum of mean square errors of parameter estimates (NLS results plotted with a solid line, the one-step method ones with a dashed line). The noise level is $\sigma=0.01,0.15, 0.25$ in the upper, middle and bottom plots, respectively. The $x$-axis is the scale parameter of the gamma distribution used to generate initial guesses for NLS, with large values corresponding to initial guesses farther away from the true parameters values.}
\end{figure}

\begin{figure}
 \centering
\includegraphics[width=0.8\textwidth,height=1.8\textheight,keepaspectratio]{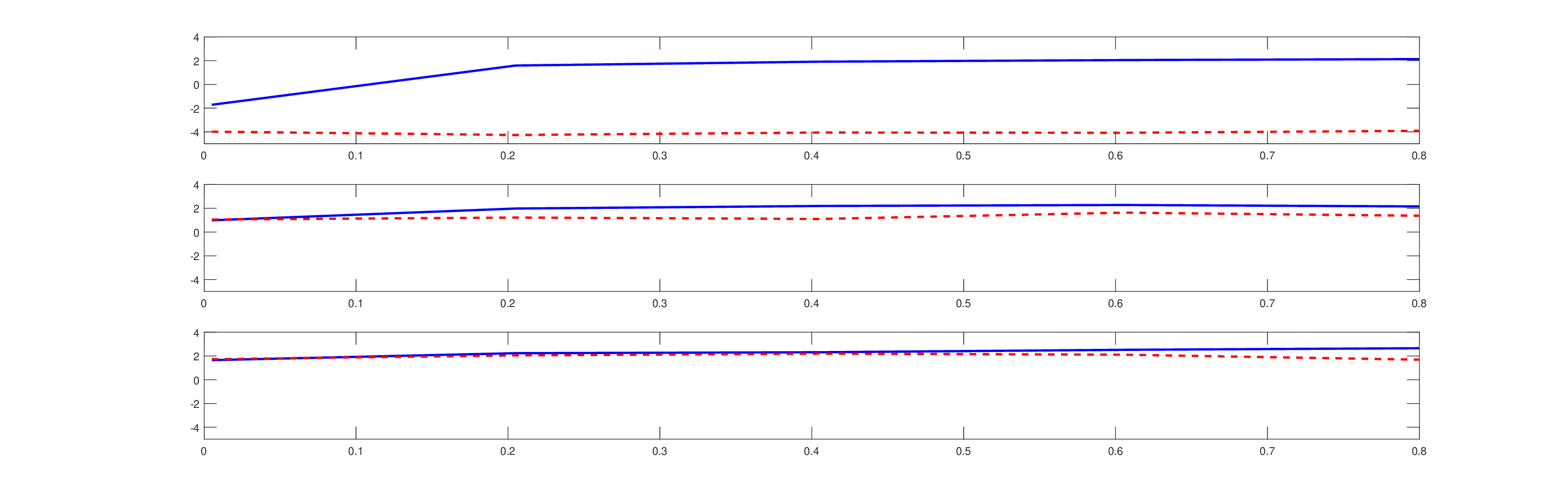}
    \caption{\label{fig:godwinSSE} Simulation results for Goodwin's oscillator in Section \ref{additional}. The $y$-axis gives the logarithm of the sum of squares of model fits (NLS results plotted with a solid line, the one-step method ones with a dashed line). The noise level is $\sigma=0.01,0.15, 0.25$ in the upper, middle and bottom plots, respectively. The $x$-axis is the scale parameter of the gamma distribution used to generate initial guesses for NLS, with large values corresponding to initial guesses farther away from the true parameters values.}
\end{figure}

We let the optimisation for NLS to start from a random initial guess generated from a gamma distribution. Specifically, the initial guess for $\theta_1$ is generated from a gamma distribution with shape parameter $\theta_1/scale$, where the scale parameter is according to the $x$-axis of Figures ~\ref{fig:godwinMSE}--\ref{fig:godwinSSE}, and similarly for $\theta_5$ the shape will be $\theta_5/scale$. The one-step method, on the other hand, employs the $\sqrt{n}$-consistent estimator, namely the direct integral estimator (although the system \eqref{goodwin} we consider is partially observed, the direct integral approach still applies, as we explain in Appendix \ref{app:goodwin}). In Figure~\ref{fig:godwinMSE} we plot on $y$-axis the logarithm of the sum of mean square errors of parameter estimates (over $100$ Monte Carlo simulation runs): NLS with a solid line, the one-step estimator with a dashed line. The noise level is $\sigma=0.01,0.15, 0.25$ in the upper, middle and bottom plots, respectively. The $x$-axis gives the scale parameter of the gamma distribution used to generate initial guesses for NLS; large values of the scale parameter correspond to a diffuse prior information on the true parameters, with initial guesses likely to be farther away from the true parameter values. In Figure~\ref{fig:godwinSSE} we show a similar setup, where now the $y$-axis gives the logarithm of the sum of squares of model fits (averaged over $100$ Monte Carlo simulation runs). 

We can see that the mean square error and the sum of squares of NLS grow together with the distance of the initial guess from the true parameter. For initial guesses close to the true parameter values, the NLS does better than the one-step method, but starts to deteriorate very quickly. Since in practice infortmative prior information on true parameters is rarely available, we conclude that the one-step method is in general better in terms of both the mean square error of parameter estimates and the sum of squares of model fits than the NLS initialised at a random initial guess. This despite the fact that we allowed the Levenberg-Marquardt implementation of NLS to run for 100 iterations, while for the one-step method we used only one iteration (as its name actually suggests). From the plots we also see that larger the measurement error, more similar the two methods are in terms of the mean square error and the sum of squares. This is not surprising, since for large noise level the direct integral estimator used as an initial input for the one-step estimator will be further away from the true parameter (as any other estimator), and hence the numerical performance of the one-step method will start to resemble that of the NLS initialised at a guess that is far from the true parameter. 

We finally remark that the pattern observed in this low-dimensional simulation example (three-dimensional system with two unknown parameters) will readily extend to the case of more complex and realistic ODE models (depending on a particular experimental setup, in an even more pronounced form).

\section{Real data examples}\label{real}

In this section we study several rea data examples.  To check the limits of applicability of the one-step method, our emphasis is on examples with small and moderate sample sizes. 

\subsection{Nitrogene oxide reaction}
The system
\begin{equation}
\label{gas}
\begin{cases}
{ x}^{\prime}(t)=\theta_1(126.2-x(t))(91.9-x(t))^2-\theta_2 (x(t))^2,\\
x(0)=0
\end{cases}
\end{equation}
describes the reversible homogeneous gas phase reaction of nitrogene oxide,
\begin{equation*}
2\rm{NO}+\rm{O}_2\rightleftharpoons 2\rm{NO}_2.
\end{equation*}
For additional chemical background see \cite{bodenstein}. Based on the experimental data from Table 39 in \cite{bodenstein}, parameters of equation \eqref{gas} were estimated via different methods in \cite{bellman1967quasilinearization}; \cite{van1975nonlinear}, see pp.\ 18--19; \cite{esposito2000global}, Section 7.4; \cite{kim2010estimation}, Section 3.1; \cite{tjoa1991simultaneous}, Problem 6 on p.\ 381; and \cite{varah1982spline}, see pp.\ 37--38. The results obtained in these papers are summarised in Table \ref{table1}.\footnote{Note that \cite{varah1982spline} gives five different parameter estimates corresponding to different values of the smoothing parameter used in his method. Of these estimates we report only the first pair and refer to Table 4 in \cite{varah1982spline} for the remaining ones. Note also that \cite{esposito2000global} use two approaches (collocation method and integration method in their terminology) and with the second of them identify another local solution to the problem, namely $\theta_1=0.1306\times 10^{-2},\theta_2=0.90393$ (see Table 11 in \cite{esposito2000global}), which we did not report in Table \ref{table1}.} We also remark that this problem is one of the six test problems in parameter estimation for ordinary differential equations that were included in \cite{floudas1999handbook}.
\begin{table}
\caption{Parameter estimates for model \eqref{gas} obtained in the literature.}
\label{table1}
\centering
\fbox{%
\begin{tabular}{l|r|r}
Paper & Estimate of $\theta_1$ & Estimate of $\theta_2$\\
\hline
\cite{bellman1967quasilinearization} & $0.4577\times 10^{-5}$ & $0.2797\times 10^{-3}$\\
\cite{van1975nonlinear} & $0.45\times 10^{-5}$ & $0.27\times 10^{-3}$\\
\cite{esposito2000global} & $0.4593 \times 10^{-5}$ & $0.28285\times 10^{-3}$\\
\cite{kim2010estimation} & $0.46\times 10^{-5}$ & $0.28\times 10^{-3}$\\
\cite{tjoa1991simultaneous} & $0.4604 \times 10^{-5}$ & $0.2847\times 10^{-3}$\\
\cite{varah1982spline} & $0.46 \times 10^{-5}$ & $0.27 \times 10^{-3}$
\end{tabular}}
\end{table}

Our interest in this example first went in the following direction: we used the realistic estimated parameter values from the literature, generated an artificial set of data from \eqref{gas} and checked how well the one-step estimator  performs in this case. We also present the estimation results using the nonlinear least squares estimator. Accordingly, we took the parameter estimates $\theta_1=0.4577\times 10^{-5}$ and $\theta_2=0.2797\times 10^{-3}$ from \cite{bellman1967quasilinearization} together with the initial condition $\xi=x(0)=0$, thus $\eta_0=(\xi,\theta_1,\theta_2)^{tr}$. Then we generated observations uniformly over $t_j\in\{0(2)40\},\ (n=21)$, according to \eqref{obs}, where the i.i.d.\ measurement errors $\epsilon_{j}$ were generated from the normal distribution $N(0,\sigma^2)$ with mean zero and variance $\sigma^2=0.25.$

This setup was chosen to mimic the real data scenario related to this model, as described later on. The fact that $\theta_1$ and $\theta_2$ are small numbers, combined with the fact that their magnitudes are rather different, renders their estimation a difficult task, cf.\ p.\ 1303 in \cite{esposito2000global}. In Table \ref{tab:nitro} we see the empirical average of point estimates and the empirical coverage of interval estimates based on Monte Carlo study consisting of $500$ runs. The point estimates are given by \eqref{opt_par}, while the interval estimates are defined in \eqref{ci}.
\begin{table}
\caption{\label{tab:nitro}
Means of point estimates and actual coverage of interval estimates for the parameters of model \eqref{gas}, where the initial value $\xi$ is zero and the parameters are $\theta_1=0.4577\times 10^{-5}$ and $\theta_2=0.2797\times 10^{-3}$. The results are based on $500$ simulation runs. There are $21$ observations given on a uniform grid on $[0,40]$, and the errors are normal with zero expectation and $\sigma^2=0.25$. The one-step point estimates are given by \eqref{opt_par}, while the interval estimates are defined in \eqref{ci}.}																				
\centering
\fbox{%
\begin{tabular}{lccccccc}
\multicolumn{3}{c}{}&
\multicolumn{2}{c}{one-step}&
\multicolumn{2}{c}{NLS}\\
\multicolumn{1}{c}{Setup}&	
\multicolumn{1}{c}{}&	
\multicolumn{1}{c}{}&
\multicolumn{1}{c}{Mean}&
\multicolumn{1}{c}{Coverage}&
\multicolumn{1}{c}{Mean}&
\multicolumn{1}{c}{Coverage}\\						
\hline	
n=21 &  $\xi_1$ &     0 & 1.491e-02 &     0.938 &7.960e-03 &    0.942  \\ 
 &  $\theta_1$ &4.577e-06 & 4.576e-06 &     0.954 &4.577e-06 &    0.952  \\ 
 &  $\theta_2$ &2.797e-04 & 2.788e-04 &     0.932 &2.798e-04 &    0.930  \\ 
\end{tabular}    }
\end{table}

We note that when estimating $\theta=(\theta_1,\theta_2),$ unlike \cite{bellman1967quasilinearization}, \cite{van1975nonlinear}, \cite{tjoa1991simultaneous} and \cite{varah1982spline}, we did not assume that the initial condition $x(0)=0$ was known, but estimated it as well. Notice also that our method exploits linearity in the parameters and therefore it is not required to supply an initial guess in the parameter space  (in \cite{bellman1967quasilinearization} and other related papers the initial guesses $\theta_1=10^{-6}$ and $\theta_2=10^{-4}$ were used). We see that even with a small sample as $21$ observations, the point and interval estimates are satisfying, and again, we do not observe a substantial difference between the one-step and NLS methods. 

We next tested our approach on the real data for the model \eqref{gas} given in Table 39 in \cite{bodenstein} and reproduced in Table I in \cite{bellman1967quasilinearization}. There are in total $14$ observations available on the interval $[0,39],$ excluding the initial condition $x(0)=0.$\footnote{Note that in Table 39 in \cite{bodenstein} and in Table I in \cite{bellman1967quasilinearization} the observation $48.8$ corresponding to the time instance $t=19$ appears to contain a typo: we tentatively corrected it to $38.8.$ The same correction was applied in Table 24 in \cite{esposito2000global} and in Table 1 in \cite{kim2010estimation}.} 
This time we did not estimate the initial condition and considered it to be zero, which agrees with the physical phenomenon the model describes. The estimation results are displayed in Table \ref{tab:nitro_real}. Both point and interval estimates obtained from the one-step and NLS methods are presented.
\begin{table}
\caption{\label{tab:nitro_real} Point estimates for the parameters of model \eqref{gas} based on the real data of Table 39 in \cite{bodenstein}. 
We consider the initial value to be zero. The one-step point estimates are given by \eqref{opt_par}; the confidence intervals were generated according to \eqref{ci}. The left and right interval points are denoted by CI(L) and CI(R), respectively.}																					
\centering
\fbox{%
\begin{tabular}{lcccccc}
\multicolumn{1}{c}{}&
\multicolumn{3}{c}{one-step}&
\multicolumn{3}{c}{NLS}\\
\multicolumn{1}{c}{}&
\multicolumn{1}{c}{Point}&
\multicolumn{1}{c}{CI(L)}&
\multicolumn{1}{c}{CI(R)}&
\multicolumn{1}{c}{Point}&
\multicolumn{1}{c}{CI(L)}&
\multicolumn{1}{c}{CI(R)}\\						
\hline	
$\theta_1$  &4.579e-06 & 4.255e-06 &  4.903e-06 &4.577e-06 &4.253e-06 &4.901e-06 \\ 
$\theta_2$  &2.791e-04 & 1.923e-04 &  3.658e-04 &2.796e-04 &1.928e-04 &3.665e-04 
\end{tabular}    }
\end{table}

\begin{figure}
 \centering
\includegraphics[width=0.45\textwidth]{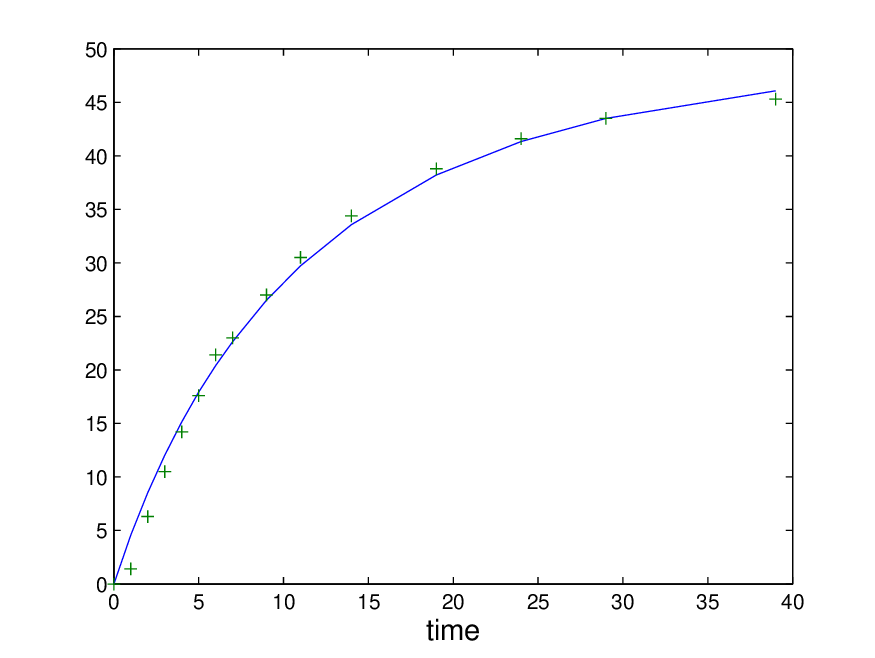}
    \caption{\label{fig:nitro_real} The solution to \eqref{gas} (given by the solid line) and the observations (indicated by pluses). The parameters were estimated using the real data from \cite{bellman1967quasilinearization}. The initial value is considered to be known and equals zero.}
\end{figure}

						
A comparison to the results given in Table \ref{table1} shows that this is essentially the same result as already reported in the literature using the least squares estimator: this illustrates the fact that one-step is an asymptotically equivalent estimator to the least squares estimator, provided a preliminary estimator it uses is already within the $n^{-1/2}$ range of the true parameter. In Figure \ref{fig:nitro_real} we plot the data from \cite{bellman1967quasilinearization} and the solution to \eqref{gas} evaluated with one-step fitted values of $\theta_1$ and $\theta_2$. The fit appears to be satisfactory given a simplistic character of the model \eqref{gas}.

\subsection{$\alpha$-pinene problem}
\begin{figure}
 \centering
\includegraphics[width=0.48\textwidth]{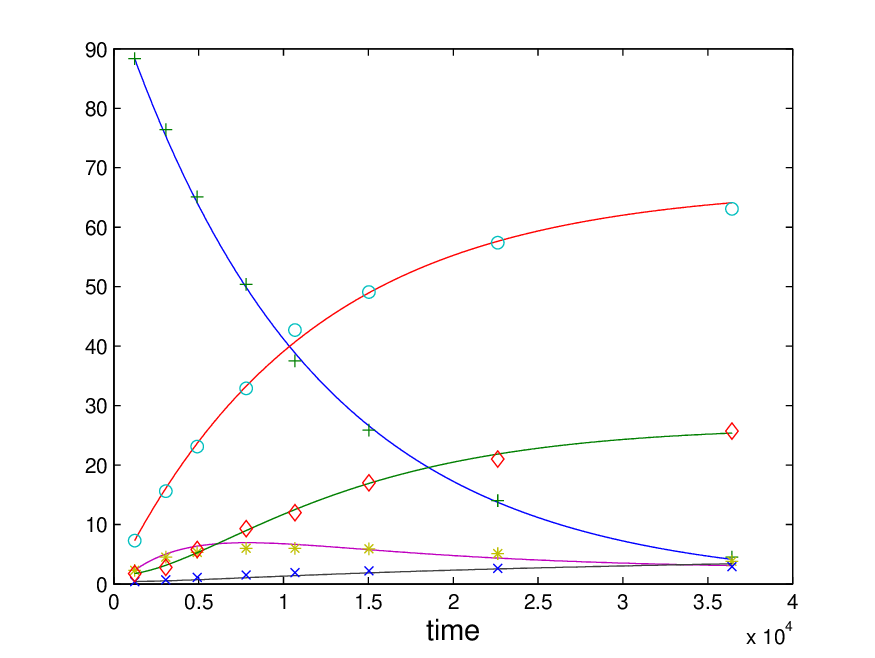}
    \caption{\label{fig:alpha_real} The solution to \eqref{alpha} based on the  one-step estimate; the observations are indicated by different symbols, corresponding to the system state they represent. The parameters were estimated using the real data from \cite{box1973some}.
 }
\end{figure}

We now consider `Problem 8' of \cite{tjoa1991simultaneous}. The system is given by 
\begin{equation}\label{alpha}
\begin{array}{l}
x_1^{\prime}(t)=-(\theta_1+\theta_2)x_1(t),
\\
x_2^{\prime}(t)=\theta_1x_1(t),
\\
x_3^{\prime}(t)=\theta_2x_1(t)-(\theta_3+\theta_4)x_3(t)+\theta_5x_5(t),
\\
x_4^{\prime}(t)=\theta_3x_3(t),
\\
x_5^{\prime}(t)=\theta_4x_3(t)-\theta_5x_5(t).
\end{array}
\end{equation}
This system characterizes a reaction that describes the thermal isomerization of $\alpha$-pinene $x_1$ to dipentene $x_2$ and alloocimene $x_3$, which in turn yields $\alpha$- and $\beta$-pyronene $x_4$ and a dimer $x_5$. The data we use are taken from Table 2 in \cite{box1973some}. For each state of the system, the data includes only $8$ observations in time. This is a challenging problem to deal with, a point raised also in \cite{tjoa1991simultaneous}, \cite{rodriguez2006} and \cite{brunel2015}. In Table \ref{tab:alpha_real} we see the resulting point and interval estimates based on the real data, using the one-step method. We do not present the results of the Monte Carlo study for the NLS method, since it could not be completed in a reasonable amount of time using the Levenberg-Marquardt method (as we did in all examples in our paper). In the last column of Table \ref{tab:alpha_real} we present the estimation result from \cite{tjoa1991simultaneous}. The solution of the system \eqref{alpha} corresponding to the one-step estimate is displayed in Figure~\ref{fig:alpha_real}. Unlike \cite{tjoa1991simultaneous}, our approach does not require to provide an initial guess in the parameter space. The parameter estimates we obtained are similar to those in \cite{tjoa1991simultaneous}, except for parameters $\theta_4,\theta_5$: the estimates computed in \cite{tjoa1991simultaneous} are not contained in our confidence intervals. As explained in detail in \cite{brunel2015}, these two parameters are the most difficult to estimate, and those authors also raise a question whether the values obtained in \cite{tjoa1991simultaneous} are reliable, and speculate the estimates in their own work could be in fact more accurate. Without offering a resolution of this difficult question, here we simply remark that alternative estimates computed in \cite{brunel2015} are contained in our confidence intervals.
\begin{table}
\caption{\label{tab:alpha_real} Point estimates for the parameters of model \eqref{alpha} based on the real data from \cite{box1973some}. 
We consider the initial values to be known. The one-step point estimates are given by \eqref{opt_par}; the confidence intervals were generated according to \eqref{ci}. The left and right interval points are denoted by CI(L) and CI(R), respectively.}																					
\centering
\fbox{%
\begin{tabular}{lcccc}
\multicolumn{1}{c}{}&
\multicolumn{1}{c}{Point}&
\multicolumn{1}{c}{CI(L)}&
\multicolumn{1}{c}{CI(R)}&
\multicolumn{1}{c}{\cite{tjoa1991simultaneous}}\\						
\hline	
$\theta_1$  &5.869e-05 & 5.771e-05 &  5.967e-05 &5.926e-05 \\ 
$\theta_2$  &2.830e-05 & 2.740e-05 &  2.920e-05 &2.963e-05 \\ 
$\theta_3$  &1.745e-05 & 1.305e-05 &  2.186e-05 &2.047e-05 \\ 
$\theta_4$  &2.132e-04 & 1.770e-04 &  2.494e-04 &2.744e-04 \\ 
$\theta_5$  &2.137e-05 & 1.037e-05 &  3.236e-05 &3.997e-05 
\end{tabular}    }
\end{table}						

Next we conducted two simulation studies, corresponding to two different measurement error variances. Specifically, we generated observations according to \eqref{obs} and \eqref{alpha} under the following experimental setup: the time grid is the same as in the real data, namely 
\[
t_j\in \{1230,3060,4920,7800,10680,15030,22620,36420\},
\]
resulting in a total of $8$ observation points. Initial values are set to the observations at the first time point, 
\[
\xi=\{88.35,7.3,2.3,0.4,1.75\}.
\]
The errors are normal with expectation zero and standard deviations 
\[\sigma=a\times\{44.6833,   36.4111,    4.9570,    1.6339,   12.4147\},
\]
corresponding to $\sigma_i$, $i=1,\ldots,5$. Here, the value $a$ is multiplied by the mean value of each state, as calculated from the solutions based on the real data example. In the first study we set $a=0.02,$ while in the second we take $a=0.1$. We note that the variance $\sigma^2$ that corresponds to $a=0.02$ is the order of the variance that we observed in the real data example. For each scenario, we repeat the experimental setup $500$ times and calculate the average of point estimates and actual coverage of the confidence intervals. We also provide the standard error of the one-step estimator as calculated based on $500$ simulations (`STE'), as well as the square root of the average of estimates of the asymptotic variance (`ASYM'). The results are presented in Table \ref{tab:alpha}. We see that the actual coverage is not too poor, but nevertheless deviates noticeably from the nominal level of $95\%.$ Further, we see a considerable difference between estimates of the asymptotic variance and the actual finite sample variance as calculated based on $500$ simulations. All these results are not surprising, if we recall that we have at hand only $8$ observations on each system state, so that asymptotic approximations are not accurate enough yet.
\begin{table}
\caption{\label{tab:alpha} Means of point estimates and actual coverage of interval estimates for the parameters of model \eqref{alpha}, where the initial value $\xi$ is considered as known. The results are based on $500$ simulation runs; see the experimental setup in the text. The one-step point estimates are given by \eqref{opt_par}; the interval estimates are defined in \eqref{ci}. Standard errors of the point estimates as calculated based on the $500$ simulations (denoted by `STE'). Square root of the average of the estimates of the asymptotic variance, over the $500$ simulations (denoted by `ASYM').}																					
\centering
\fbox{%
\begin{tabular}{lcccccc}
\multicolumn{1}{c}{Setup}&
\multicolumn{1}{c}{}&
\multicolumn{1}{c}{True}&	
\multicolumn{1}{c}{Mean}&
\multicolumn{1}{c}{Coverage}&
\multicolumn{1}{c}{STE}&
\multicolumn{1}{c}{ASYM}\\			
\hline	
$\sigma=0.02$ &  $\theta_1$ &5.926e-05 & 5.920e-05 &     0.758 &  6.539e-07 &  3.913e-07\\ 
 &  $\theta_2$ &2.963e-05 & 2.958e-05 &     0.806 &  5.246e-07 &  3.615e-07\\ 
 &  $\theta_3$ &2.047e-05 & 2.042e-05 &     1.000 &  5.789e-07 &  1.815e-06\\ 
 &  $\theta_4$ &2.744e-04 & 2.709e-04 &     1.000 &  7.847e-06 &  2.099e-05\\ 
 &  $\theta_5$ &3.997e-05 & 3.878e-05 &     0.998 &  2.793e-06 &  6.060e-06\\ 
\hline
$\sigma=0.1$ &  $\theta_1$ &5.926e-05 & 5.910e-05 &     0.768 &  3.265e-06 &  3.026e-02\\ 
 &  $\theta_2$ &2.963e-05 & 2.945e-05 &     0.820 &  2.669e-06 &  2.717e-03\\ 
 &  $\theta_3$ &2.047e-05 & 1.993e-05 &     0.998 &  2.755e-06 &  9.746e-06\\ 
 &  $\theta_4$ &2.744e-04 & 2.452e-04 &     0.946 &  8.382e-05 &  1.406e-04\\ 
 &  $\theta_5$ &3.997e-05 & 3.103e-05 &     0.940 &  2.569e-05 &  9.688e-05 
\end{tabular}    }
\end{table}						

\section{Conclusions}\label{sum}

Parameter estimation for ODEs is a challenging problem. In this paper we have explored performance of Le Cam's one-step method in the ODE context both from applied and theoretical sides. Using real and simulated data examples, we have demonstrated that execution of a one-step correction on a preliminary smoothing-based estimator leads to rather satisfactory estimation results, that are comparable to those in the `gold standard' least squares estimation. In particular, we can argue that already for small and moderate sample sizes the one-step method yields results comparable to the nonlinear least squares estimation in terms of the statistical accuracy, as suggested by the asymptotic statistical theory. The empirical coverage of the confidence intervals we provide is good even for samples as small as $n=21$ in the examples we considered. On the other hand, for very small sample sizes the nonlinear least squares method appears to perform better than the one-step method, though the latter remains reasonable. Furthermore, we note that the one-step approach discussed in this work was applied for both fully and partially observed ODE systems (see Section \ref{additional}). 

The relation between the one-step method and the Levenberg-Marquardt method we pointed out in Section \ref{additional} leads to a very simple practical implementation: when computational time is an issue, our simulations and theory justify the use of the Levenberg-Marquardt algorithm with only one iteration, if its starting point is SME or the integral estimator. In particular, as evidenced by the results presented in Section \ref{additional}, the performance of the one-step estimator  is as good as or even  better than that of the NLS starting from a random initial guess and using 100 iterations. This is a useful practical observation: tuning the number of iterations is possible in software implementations of optimisation algorithms, such as the one in Matlab, and hence the one-step correction on the SME or the integral estimator is straightforward to implement. 




\section*{Acknowledgements}
The idea of using the one-step Le Cam method in the context of parameter inference for ODEs was proposed to us by C.A.J. Klaassen (University of Amsterdam), who was also involved in early stages of the present research. We would like to thank him for most stimulating discussions and helpful remarks.

The first author was supported by the Israeli Science Foundation grant number 387/15, and by a Grant from the GIF, the German-Israeli Foundation for Scientific Research and Development number I-2390-304.6/2015. The second author was supported by the European Research Council under ERC Grant Agreement 320637.

\appendix
\section{Proof of Theorem 1}
\label{appA}
Note that $\overline{\eta}_n^{\ast}=\overline{\eta}_n(\hat{\eta}_{\overline{\rho}_n})$ for some data-dependent (random) smoothing parameter $\overline{\rho}_n$ taking values in the set $\mathcal{R}_n$; more formally,
\begin{equation*}
\overline{\rho}_n=\operatorname{argmin}_{ \rho_n(k)\in R_n } \sum_{i=1}^d\sum_{j=1}^n(Y_{ij}-x_i(\overline{\eta}(\hat{\eta}_{\rho_n(k)}),t_j))^2.
\end{equation*}
Observe that the estimator $\hat{\eta}_{\overline{\rho}_n}$ is $\sqrt{n}$-consistent. This claim appears to be self-evident, but nevertheless, we still provide its proof. Thus, for every fixed $\varepsilon>0,$ we have to show existence of a constant $K_{\varepsilon},$ such that
\begin{equation*}
P( \sqrt{n} | \hat{\eta}_{\overline{\rho}_n} - \eta_0  | \geq K_{\varepsilon} )\leq \varepsilon
\end{equation*}
for all $n\geq n_{\varepsilon},$ where $n_{\varepsilon}$ is some integer, possibly depending on $\varepsilon$ and $K_{\varepsilon}.$ We have
\begin{align*}
P( \sqrt{n} | \hat{\eta}_{\overline{\rho}_n} - \eta_0  | \geq K_{\varepsilon} ) & \leq P\left( \sqrt{n} \sum_{i=1}^N| \hat{\eta}_{{\rho}_n(i)} - \eta_0  | \geq K_{\varepsilon} \right)\\
& \leq \sum_{i=1}^N P\left( \sqrt{n} | \hat{\eta}_{{\rho}_n(i)} - \eta_0  | \geq \frac{ K_{\varepsilon}}{N} \right).
\end{align*}
$\sqrt{n}$-consistency of $\hat{\eta}_{\overline{\rho}_n}$ now easily follows from the above inequality and $\sqrt{n}$-consistency of each $\hat{\eta}_{{\rho}_n(k)},$ $k=1,\ldots,N.$

Now that we know the estimator $\hat{\eta}_{\overline{\rho}_n}$ is $\sqrt{n}$-consistent, the proof of our theorem consists in application of Theorem 5.45 and Addendum 5.46 in van der Vaart (1998), which in turn can be reduced to verification of conditions of Theorem 5.41 there. This amounts to verification of the following conditions:
\begin{enumerate}
\item It must hold that $\sqrt{n}\Psi_n(\eta_0)$ converges in distribution. Here $\Psi_n$ is as in formula \eqref{esteq}.
\item It must hold that for every fixed $(t,y),$ the function $\psi_{\eta}(t,y)$ is twice continuously differentiable with respect to $\eta.$ Here $\psi_{\eta}(t,y)$ is as in \eqref{psieta}.
\item It must hold that $\ex[\psi_{\eta_0}(T_1,Y_1)]=0,$ $\ex[| \psi_{\eta_0}(T_1,Y_1) |^2]<\infty,$ and the matrix
\begin{equation*}
\ex\left[ \frac{d}{d\eta}\psi_{\eta} (T_1,Y_1)|_{\eta=\eta_0} \right]
\end{equation*}
must be nonsingular. Here $Y_1$ is a shorthand notation for the vector $(Y_{11},Y_{21},\ldots,Y_{d1}).$
\item It must hold that the second order partial derivatives of the function $\psi_{\eta}$ with respect to $\eta_j,\eta_k$ are dominated by an integrable (with respect to its distribution) function of $(T_1,Y_1).$
\end{enumerate}

Arguments for verification of these conditions are quite standard and follow from the regularity assumptions in the statement of our theorem. The limit covariance matrix in \eqref{onestep} is obtained in the process of verification of (i)-(iv) above.

\section{Integral estimator}
\label{appB}

Given observations $Y_{ij}$'s, the one-step method requires first to have at hand a $\sqrt{n}$-consistent estimator of $\theta_0$ and $\xi_0$. As mentioned in the previous sections, the SME provides us with such an estimator. However, this method is based on estimating the derivative $x^\prime,$ which is hard to do accurately in practice for small or moderate sample sizes. In the case where the symbol $F$ of the system of ODEs is linear in functions of the parameter $\theta$, one can avoid estimation of derivatives and use an integral SME. Indeed, in such cases one can use some version of the so called 'integral approach' (see \cite{himmelblau1967determination}) as was studied in \cite{dattner2015}. The idea works as follows: note that for systems whose symbols are linear in parameters, $F(x(t);\theta)=g(x(t))\theta$
holds, where the measurable function $g:\rR^d \to \rR^{d\times p}$ maps
the $d$-dimensional column vector $x$ into a $d\times p$ matrix. Let $\hat{x}_n(\cdot)$ be an estimator of $x(\eta_0,\cdot)$, and denote $\hat{G}_n(t)=\int_0^t g(\hat{x}_n(s),s)\rd s$, $\hat{A}_n=\int_0^T \hat{G}_n(t)\rd t$, $\hat{B}_n=\int_0^T\hat{G}_n^T(t)\hat{G}_n(t)\rd t$, and let $I_d$ be the $d \times d$ identity matrix. Then \cite{dattner2015} show that the direct estimators
\begin{eqnarray}
\hat{\xi}_n &=& \left(I_d - \hat{A}_n \hat{B}_n^{-1}
\hat{A}_n^T\right)^{-1} \int_0^T \left(I_d - \hat{A}_n
\hat{B}_n^{-1}
\hat{G}_n^T(t)\right) \hat{x}_n(t)\,\rd t, \label{xihat} \\
\hat{\theta}_n &=& \hat{B}_n^{-1} \int_0^T \hat{G}_n^T(t) \left(
\hat{x}_n(t) -\hat{\xi}_n \right) \rd t, \label{thetahat}
\end{eqnarray}
are $\sqrt{n}$-consistent. In case the initial value $\xi_0$ is known, (\ref{thetahat}) may be used with $\hat{\xi}_n$ replaced by $\xi_0$. Besides the required statistical properties, the extensive simulation study presented in the aforementioned paper suggests that this approach is much more accurate in finite samples compared to the derivative-based SME. Thus, we use the integral SME \eqref{xihat}--\eqref{thetahat} whenever applicable, and the derivative-based SME otherwise.  

We choose to estimate the solution $x$ using local polynomial estimators, which are consistent and `automatically' correct for the boundaries. Under the assumption that $x$ are $C^\alpha$-functions for some real $\alpha \geq 1,$ we will approximate them by polynomials of degree
$\ell=\lfloor\alpha\rfloor$ as follows (Tsybakov (2009), Section 1.6): let
\begin{eqnarray*}\label{LPestimator}
U(u)&=&\Big(1,u,u^2/(2!),...,u^\ell/(\ell!)\Big)^T,\quad u\in \mathbb R,\\
\nu(t)&=&\left(x(t),x^\prime(t)b,x^{\prime\prime}(t)b^2,...,x^{(\ell)}(t)b^\ell\right),\quad
t \in \mathbb R, \nonumber
\end{eqnarray*}
where $b=b_n>0$ is a bandwidth, the $(\ell+1)$-vector $U(u)$ is a
column vector, and $\nu(t)$ is a $d \times (\ell+1)$-matrix. Let
$K(\cdot)$ be some appropriate kernel function and define
\begin{eqnarray*}
\hat{\nu}_n(t)&=&\arg\min_{\nu\in\rR^{d \times (\ell+1)}}
\sum_{i=1}^n \Big\{Y(t_i)-\nu U\Big(\frac{t_i-t}{b}\Big)\Big\}^T
\\&&\times\Big\{Y(t_i)-\nu U\Big(\frac{t_i-t}{b}\Big)\Big\}
K\Big(\frac{t_i-t}{b}\Big).
\end{eqnarray*}
The local polynomial estimator of order $\ell$ of $x(t)$ is the first column of the $d\times (\ell+1)$-matrix $\hat{\nu}_n(t)$, i.e., $\hat{x}_n(t)=\hat{\nu}_n(t)U(0)$. 

We applied the estimation procedure described above to a set of bandwidths $B:=\{b_{\min},\ldots,b_{\max}\}$, and for a given $b\in B$ we denote the resulting one-step parameter estimator by $\bar\eta_{n,b}$. We then select $\bar\eta_n=\bar\eta_{n,\bar b}$ for some $\bar b\in B,$ the choice of which is discussed in Remark \ref{consistency_sme_rem} of the main text. Last, we use local estimators polynomials of order $1$, with $K(t)=3/4(1-t^2)\textbf{1}\{|t|\leq 1\}$ (cf.\ \cite{dattner2015}), where $\textbf{1}\{\cdot\}$ stands for the indicator function. Other kernels are also possible.

\section{Goodwin's oscillator}
\label{app:goodwin}

In Section \ref{additional} we applied the direct integral method on a partially observed Goodwin's oscillator,
\begin{equation}
\label{goodwin2}
\begin{split}
\begin{cases}
{ x}^{\prime}_1(t)&=\frac{\theta_1}{1+2x_3(t)^{10}}-\theta_5 x_1(t),\\
{ x}^{\prime}_2(t)&=2 x_1(t)-\theta_5 x_2(t),\\
{ x}^{\prime}_3(t)&= x_2(t)-\theta_5 x_3(t).
\end{cases}
\end{split}
\end{equation}
The integral estimation approach works as follows in this case: first apply the integral estimation method from Appendix \ref{appB} on the second equation of \eqref{goodwin2} and obtain a $\sqrt{n}$-consistent estimator of $\theta_5$ (this is possible, because the state variable $x_2$ is observed in the setting of Section \ref{additional}). Next integrate the equation
\begin{equation*}
\begin{split}
\begin{cases}
{ x}^{\prime}_3(t)&= \hat{x}_2(t)-\hat{\theta}_5 x_3(t),\\
x_3(0)&=0,
\end{cases}
\end{split}
\end{equation*}
to get an estimator $\hat{x}_3$ of the component $x_3$ of the solution to \eqref{goodwin2}. Finally, apply the integral estimation method on the first equation of \eqref{goodwin2} to get a $\sqrt{n}$-consistent estimator of $\theta_1$ (this is possible, since estimators $\hat{x}_1$ and $\hat{x}_3$ of $x_1$ and $x_3$ are available, the first one because the variable $x_1$ is observable in the setting of Section \ref{additional}).

\bibliographystyle{chicago}
\bibliography{bib}

\end{document}